\documentclass{aa}
\usepackage{graphicx}
\usepackage{txfonts}
\usepackage{natbib,twoopt}
\bibpunct{(}{)}{;}{a}{}{,}    

\usepackage[dvipsnames]{xcolor}
\usepackage{makecell}
\usepackage{hyperref} 
\hypersetup{
  colorlinks = true, 
  urlcolor   = blue, 
  linkcolor  = blue, 
  citecolor  = blue 
}
\newcommand{\MgI}{\ion{Mg}{I}\xspace}
\newcommand{\MgII}{\ion{Mg}{II}\xspace}
\newcommand{\MgIII}{\ion{Mg}{III}\xspace}
\newcommand{\NiI}{\ion{Ni}{I}\xspace}

\makeatletter
\renewcommand*\aa@pageof{, page \thepage{} of \pageref*{LastPage}}
\makeatother

\begin{document} 
   
\title{Modeling the \MgI from the NUV to MIR: I. The Solar Case}

\author{
    J.I. Peralta\inst{1,2}
    \and Vieytes M.C.\inst{1,2}
    \and A.M.P. Mendez\inst{1}
    \and D.M. Mitnik \inst{1,3}
}

\institute{Instituto de Astronomía y Física del Espacio, CONICET--Universidad de Buenos Aires, Argentina\\
        \email{jperalta@iafe.uba.ar; mariela@iafe.uba.ar}
        \and Departamento de Ciencia y Tecnología, UNTREF, Argentina
        \and Departamento de Física, Universidad de Buenos Aires, Argentina
}

\date{Received 6 August 2021, Accepted 6 October 2021}

\abstract
{
    Semi-empirical models of the solar atmosphere are used to study the radiative environment of any planet in our solar system. There is a need for reliable atomic data for neutral atoms and ions in the atmosphere to obtain improved calculated spectra. Atomic parameters are crucial to computing the correct population of elements through the whole stellar atmosphere. Although there is a very good agreement between the observed and calculated spectra for the Sun, there is still a mismatch in several spectral ranges due to the lack of atomic data and its inaccuracies, particularly for neutrals like \MgI. 
}
{
    To correctly represent many spectral lines of \MgI from the near-UV to the mid-IR is necessary to add and update the atomic data involved in the atomic processes that drive their formation. 
}
{
    The improvements to the \MgI atomic model are as follows: 
    \textit{i)} 127 strong lines, including their broadening data, were added.
    \textit{ii)} To obtain these new lines, we increased from 26 to 85 the number of energy levels.
    \textit{iii)} Photoionization cross-section parameters were added and updated.
    \textit{iv)} Effective Collision Strengths $(\Upsilon_{ij})$ parameters  were updated for the first 25 levels using the existing data from the convergent close-coupling (CCC) calculations. One of the most significant changes in our model is given by the new $\Upsilon_{ij}$ parameters for transitions involving levels between 26 and 54, which were computed with a multi–configuration Breit–Pauli distorted–wave (DW) method. For transitions involving superlevels, we calculated the $\Upsilon_{ij}$ parameters with the usual semi-empirical formulas of \citet{seaton:1962} and \citet{vanRegemorter:1962}.
}
{
    More than one hundred transitions were added to our calculations, increasing our capability of reproducing important features observed in the solar spectra. We found a remarkable improvement in matching the solar spectra for wavelengths higher than 30\,000 {\AA} when our new DW $\Upsilon_{ij}$ data was used in the model.  
}
 

\maketitle

\section{Introduction} \label{sec:intro}

1D semi-empirical model calculation of the solar atmosphere continues to be useful to predict and reproduce the Solar Spectral Irradiance (SSI), from the far-infrared (FIR) to the extreme ultraviolet (EUV), received at the top of the Earth's atmosphere \citep{fontenla:2018}. Changes in the SSI can produce modifications in the chemistry of our atmosphere. 

The advantage of these models is that they are built to reproduce a large number of observations in different wavelengths. The equations of statistical equilibrium and radiative transfer are treated in detail and solved simultaneously for all the species in the atmosphere.  
One of the most critical parameters is the data describing the atomic and molecular species present in the atmosphere. Great efforts have been made to obtain experimental and theoretical data. However, there is still a lack of complete and accurate atomic data, particularly for neutral and low ionized atoms. These atomic species, such as \MgI, are of particular interest because they have relevant spectral features used in the diagnosis of solar and stellar astrophysics. 
The influence of using a more complete and accurate atomic model for neutral atoms in SSI calculations can be found, for example, in \cite{vieytes:2013}, were the improvements made in the \NiI atomic model led to a better match with observations, mainly in the near-ultraviolet (NUV) region. 

Neutral magnesium is an important chemical element since its lines are strong in the spectra of late-type and even in metal-poor stars, making it a good tracer of $\alpha$-element abundances. At chromospheric temperatures, magnesium is susceptible to deviations from local thermodynamic equilibrium (LTE). These deviations from LTE are predicted to be significant, particularly for metal-poor stars \citep{zhao:1998,zhao:2000}. Therefore to study the non-LTE (NLTE) effects it is necessary to account for accurate atomic data. 

Past studies have used semi-empirical formulae when computing electron atomic collision data. The most widely used semi-classical formulas for radiative allowed transitions are the ones given by \cite{seaton:1962} and \cite{vanRegemorter:1962} since they provide estimates for the collision rates based on the oscillator strengths and transition energies. However, these approximations can lead to a great number of uncertainties in the synthetic spectra of NLTE atmospheric models, such as spectral lines in emission instead of in absorption as observations reveal. 

There are several studies on NLTE calculations of \MgI in the Sun and other stars, for instance, the recent works by \cite{alexeeva:2018}, the paper series by \cite{osorio:2015}, \cite{osorio:2016} and \cite{barklem:2017}, and the relevant works by \cite{bergemann:2015} and \cite{scott:2015}. In all of these studies, the authors stress the need for accurate \MgI atomic data.

\cite{osorio:2015} described in detail previous studies of NLTE \MgI line formation. They presented a complete \MgI atomic model of 108 energetic states (including fine-structure splitting), up to level $20d$. Among other important improvements, they included new electron impact values for low-lying states using the R-Matrix method. For transitions involving states greater than $5p\,^3$P the data was complemented with the \cite{seaton:1962} formula. The authors analyzed relevant spectral lines from $\sim$3800 {\AA} to $\sim$8800 {\AA} including the mid-IR (MIR) lines: 7.3, 12.2, 12.3, 18.8, 18.96 $\mu$m. The NLTE calculation was computed using the \textit{MULTI} code \citep{carlsson:1992}. Their synthetic spectra showed a good agreement with solar observations and with observations of five F-G-K late-type stars with reliable fundamental parameters. \cite{barklem:2017} presented the latest calculations for inelastic e+Mg for the first 25 energy states of neutral magnesium (up to $3s6p\,^1P$), produced by the convergent close-coupling (CCC) and the B-spline R-matrix (BSR) methods. In this work, the authors suggest their CCC data for NLTE modeling.

Later, \cite{alexeeva:2018} made an extensive study of NLTE \MgI line formation in the Sun and 17 stars, covering B-A-F-G-K spectral types. They used an atomic model of about 89 energetic states, up to level $n=16$. For electronic collisions, they implemented data from \cite{osorio:2015} and complemented with data from \cite{seaton:1962} formula for allowed transitions. In the case of forbidden transitions, the authors assumed a single value for the effective collision strength, being $\Upsilon_{ij}=1$. To test their results they also selected spectral lines starting at $\sim$3800 {\AA} until $\sim$8800 {\AA}, and included the MIR lines: 7.3, 12.2, 12.3 $\mu$m. The NLTE calculation was made using the \textit{DETAIL} code \citep{butler:1985}.

None of the previous works on \MgI line formation reproduces spectral lines in the NUV region or lines between $\sim$8800 {\AA} and 7.3 $\mu$m in the NIR and MIR. Nevertheless, the NUV plays a fundamental role in ozone formation, and any calculation of the SSI must reproduce this region as well as possible. 

Everything we know about our Sun and its relationship with the Earth's atmosphere can be extended and applied to study the atmosphere of exoplanets orbiting around late-type stars. The same methodology is used to calculate the Spectral Energy Distribution (SED) that an exoplanet orbiting a late-type star receives at the top of its atmosphere \citep{tilipman:2020}. 

Nonetheless, our Sun can be observed with unique spatial and spectral resolution. For this reason, changes in the atomic models included in the atmospheric modeling must first be tested to reproduce the observed solar spectra.

The goal of the present work is two folded. We intend to update the \MgI atomic model. Consequently, we expect to extend the analysis of spectral line formation of \MgI to include transitions in a broader range than previous studies, from the NUV (1700 {\AA}) to MIR (72\,000 {\AA}).

Regarding the electronic collisions, we included two significant improvements to the electron impact excitation data. From our original atomic model, we replaced the semi-empirical effective collision strength $(\Upsilon_{ij})$ values for the first 25 lower-lying terms by close-coupling calculations \citep{barklem:2017}. Furthermore, we used the \textsc{autostructure} code \citep{badnell:2011} to calculate collisional strengths under the distorted wave approximation. We incorporated these results by replacing the semi-empirical $\Upsilon_{ij}$ involving states between $3s6p$ and $3s7i$.

This paper is structured as follows: The code used in our NLTE and spectra calculations together with the initial atomic model is detailed in Section \ref{sec:NLTE_calculations}. Section \ref{sec:new_model} shows the different \MgI atomic models built. The observations used to compare with our synthetic spectra are described in Section \ref{sec:observations}.  Our results and discussion are presented in Section \ref{sec:results}, while our final remarks and conclusions are shown in Section \ref{sec:conclusions}.

\begin{figure}
\centering
\resizebox{\hsize}{!}{\includegraphics{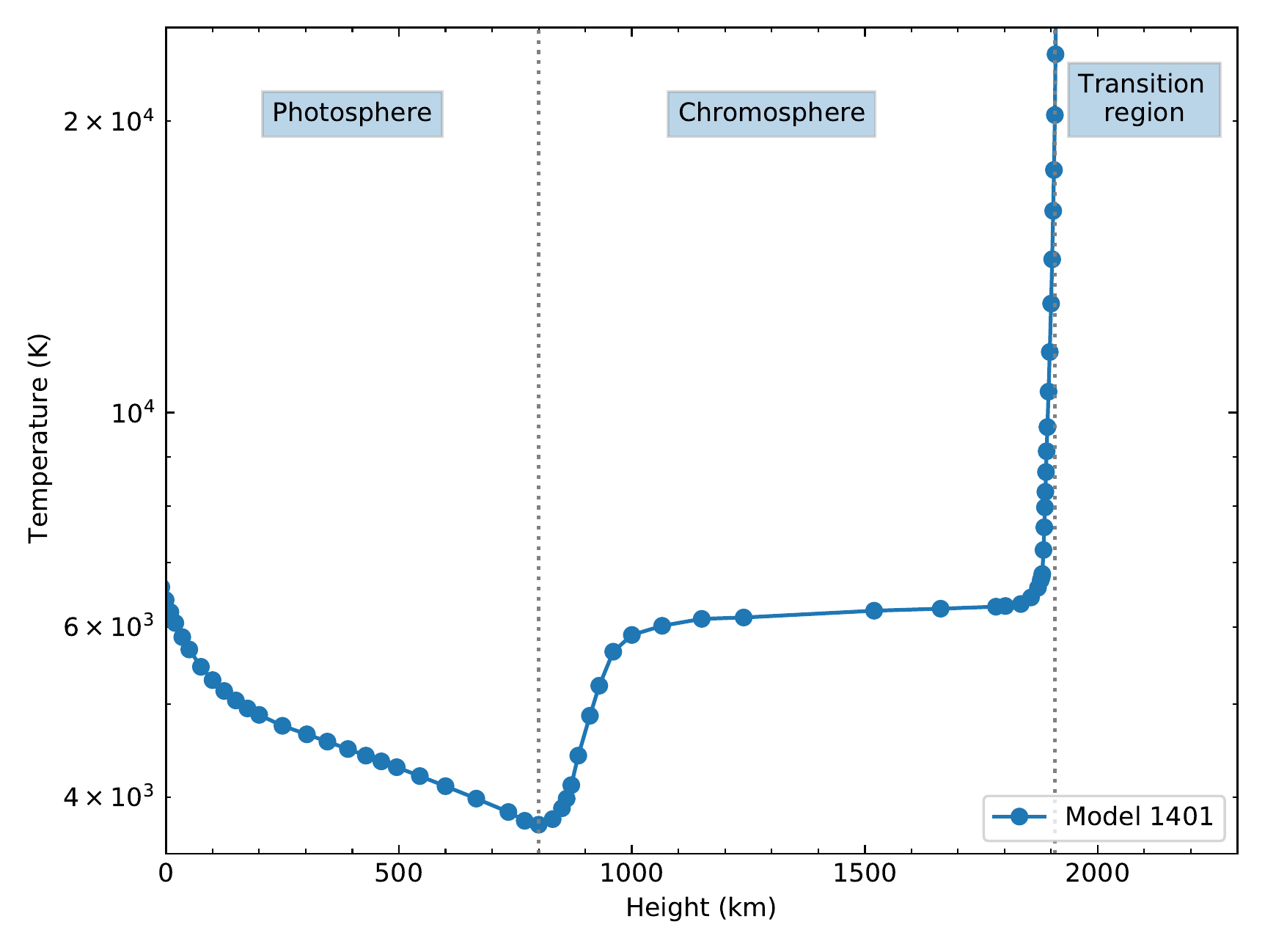}}
\caption{Atmospheric model 1401 by \cite{fontenla:2015}. Models in this work have the same thermal structure than 1401.}
\label{fig:1401}
\end{figure}

\section{NLTE and initial emergent spectra calculations} \label{sec:NLTE_calculations}

\begin{figure}
\centering
\resizebox{\hsize}{!}{\includegraphics{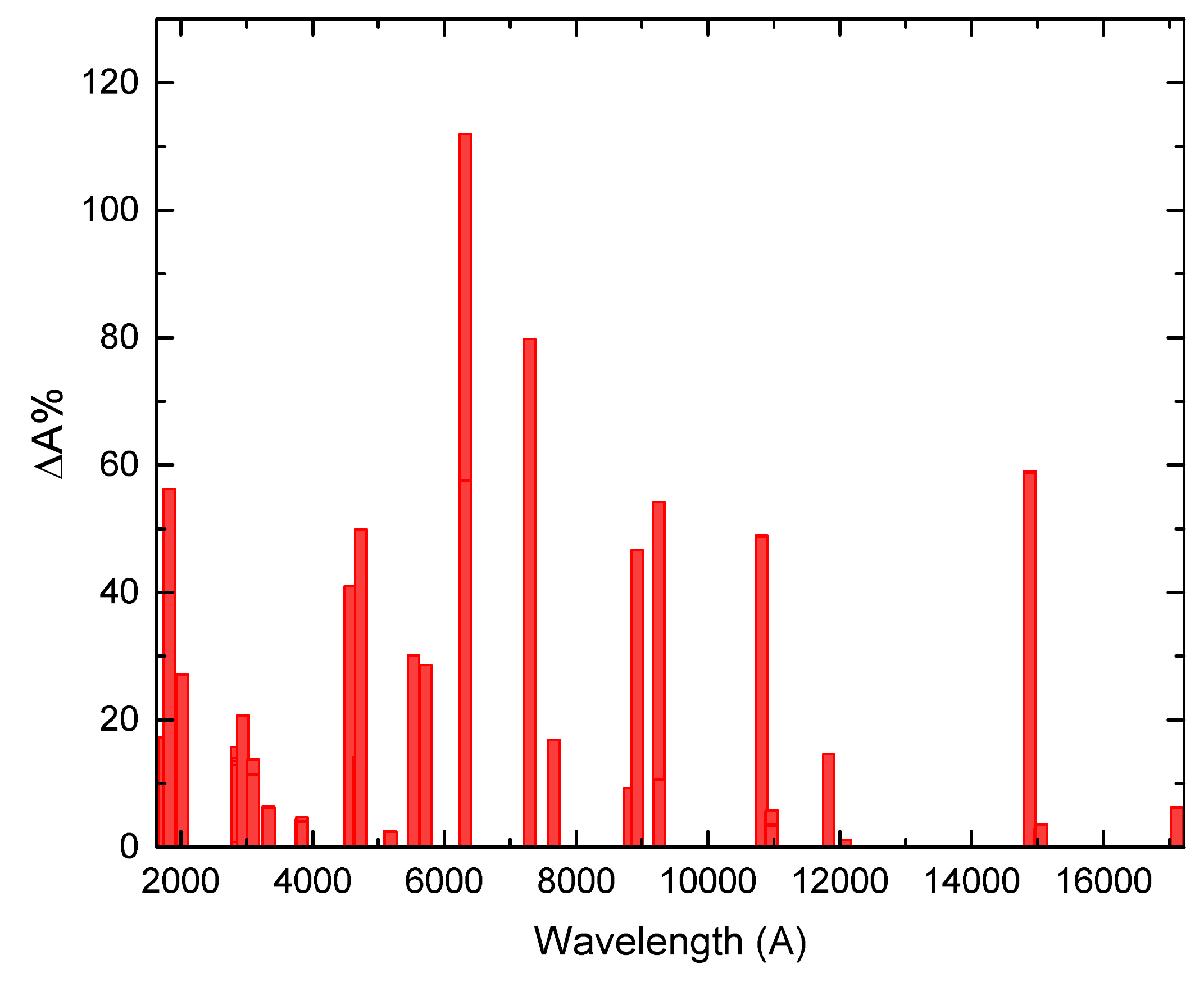}}
\caption{Distribution of changes between the original and the updated \textit{A} coefficients with wavelength included in model 1401a.}
\label{fig:achanges}
\end{figure}

\begin{table*}
\centering
\begin{tabular}{||l|c|c|l|l||}
 \hline
Model & \MgI levs & \MgII levs & \multicolumn{1}{c|}{$\Upsilon_{ij}$ methods (\MgI) } & \multicolumn{1}{c||}{Notes}\\ 
 \hline\hline
 1401  & 26 & 14 & SEA\&VRM & Original model in \cite{fontenla:2015} \\ \hline
 1401a & 26 & 14 & SEA\&VRM & \textit{gf} and broad. data updated \\ \hline
 1401b & 85 & 47 & CCC (< lev 26) + SEA\&VRM (levs 26 to 85) & \textit{gf}, broad. and photoioniz. data updated \\ \hline
 1401c & 85 & 47 & \begin{tabular}{@{}l@{}}
                    CCC (< lev 26) + DW (levs 26 to 54) +\\
                    SEA\&VRM (levs 55 to 85)
                    \end{tabular} & \textit{gf}, broad. and photoioniz. data updated \\ 
\hline
\end{tabular}
\caption{Models summary. The atmospheric structure of model 1401 \citep{fontenla:2015} is used in each model built.} 
\label{table:models}
\end{table*}

\subsection{\textit{SRPM} system and the solar model employed} \label{subsec:SRPM_sun}

We carried out the NLTE and spectra calculations by making use of the \textit{Solar Radiation Physical Modeling (SRPM)} version 2 system, a first version developed by \cite{fontenla:2005}, and later updated by \cite{fontenla:2015}. This system is a library of codes that allows us to calculate self-consistently the equations of statistical equilibrium and radiative transfer for a plane-parallel or spherical symmetric atmosphere, assuming hydrostatic equilibrium.

One key point of the \textit{SRPM} system is that it computes full NLTE for all atomic species in the atmosphere, based on the Net Radiative Bracket Operator formulation implemented by \cite{fontenlarovira_a:1985}. This method was further developed to include Partial Redistribution (PRD) \citep{fontenla:1996}, particle diffusion and flows, as outlined by \cite{fontenla:1990} and \cite{fontenla:1991}. 

The \textit{SRPM} calculates the populations of $H$, $H^{-}$, $H_{2}$ and 52 neutral and lowly ionized atomic species (see Table 2 in \cite{fontenla:2015}) in optically thick full NLTE and PRD. Additionally, the optically thin NLTE approximation is used for 198 highly ionized species. Our atomic database takes into account 18\,538 levels and 435\,986 transitions produced by atoms and ions. Apart from atomic species, the 20 most abundant diatomic molecules and over 2 million molecular lines are included.
 
The \textit{SRPM} has been widely used to build models for each different feature observed on the solar disk, and then successfully tested against ground and space-based solar spectra. These models can accurately reproduce observations of the Solar Spectral Irradiance (SSI), including the UV region observed by a number of space missions with good absolute flux calibration (see \cite{fontenla:2018} and references therein for more detail).

The solar atmospheric model used to calculate the NLTE populations and the solar spectra for the different sets of \MgI atomic models is the model 1401 built by \cite{fontenla:2015}, which is presented in Fig. \ref{fig:1401}. This model reproduces the quiet sun inter-network (see Table 1 in that paper), and represents the different layers of the solar atmosphere, the photosphere, the chromosphere, and the transition region, as shown in Fig. \ref{fig:1401}. Considering solar disk observations in the minimum of the activity cycle, the quiet sun inter-network is the dominant feature on the disk when a multi-component model for the atmosphere of the quiet Sun is calculated.

\subsection{Initial atomic model} \label{subsec:initial_model}

We describe in this section the initial atomic models for the neutral and the first two ionized atoms, \MgI, \MgII, and \MgIII. These atomic models constitute the starting model 1401. The population of these species was calculated in full NLTE. It is important to note that we also considered the rest of the ionization states in NLTE optically thin, and they were taken into account for the ionization balance of the Mg atomic element. 

\textbf{\MgI}. The original atomic model was developed by \cite{fontenla:2006} with data from \textit{NIST}\footnote{\url{https://www.nist.gov/pml/atomic-spectra-database}} \citep{nist:2004}. This model included 26 energy levels (up to $3s5g\,^3G$, 57\,262.76 $cm^{-1}$) with their fine-structure splitting, which produce a total of 44 energy sublevels\footnote{Clarification: we follow the common nomenclature: ``level'', to refer to $^{2S+1}L$ term, and ``sublevel'' when referring to a fine-structure $^{2S+1}L_J$ level.}.
This model included 82 \MgI spectral lines in the range of 1700--18\,000 {\AA}.
The $\Upsilon_{ij}$ parameters for excitation due collision with electrons were obtained from the semi-empirical equations of \cite{seaton:1962} (hereafter SEA) for allowed transitions and  \cite{vanRegemorter:1962} (hereafter VRM) when transitions were forbidden. For ionizing collisions, the \cite{NRL:2005} ionization rate equation was used.
Photoionization cross-sections were taken from \textit{TOPbase}\footnote{\url{http://cdsweb.u-strasbg.fr/topbase/xsections.html}} the Opacity Project atomic database \citep{topbase:1993}. \MgI data included cross-sections up to level $3s5g\,^{1,3}G$ (level 26 in our database). The radiative recombination is included as the inverse of the photoionization process. We also considered the dielectronic recombination for the ionization balance, but only for the ground state of the species. These values were taken from \textit{CHIANTI} version 7.1 \citep{chianti:2012}.
For the line broadening parameters concerning radiative, Stark, and van der Waals processes, the approximate values by \cite{kurucz:1995} were used. 

\textbf{\MgII}. The atomic model included 14 levels (up to $6p\,^2P$, 105\,622.34 $cm^{-1}$), with a total of 23 energy states considering the fine-structure splitting. With this structure, 52 line-transitions were included (22 for term-term transitions) in the range of 1020--11\,000 {\AA}. The $\Upsilon_{ij}$ data were extracted from \textit{CHIANTI}\footnote{\url{https://www.chiantidatabase.org/}} atomic database (versions 5.2 and 7.1) and complemented with SEA (allowed transitions) and VRM (forbidden transitions) when missing in \textit{CHIANTI}. For ionizing collisions, the NRL equation was used. \MgII photoionization cross-sections data was taken from \textit{TOPbase}. Radiative and dielectronic recombination were considered as in \MgI.

\textbf{\MgIII}. The model included 54 energy levels (up to $2s^2\,2p^5$ $[^2P_{3/2}]\,6d$, 618\,601 $cm^{-1}$) and 346 spectral lines (141 term-term transitions). For $\Upsilon_{ij}$ parameters, the SEA\&VRM combination is used. \MgIII photoionization cross-sections parameters from \textit{TOPbase} for all levels in the database are included. Radiative and dielectronic recombination were considered as in the previous ions. The \MgIII atomic model remains unchanged during this work.

\begin{figure}
\centering
\resizebox{\hsize}{!}{\includegraphics{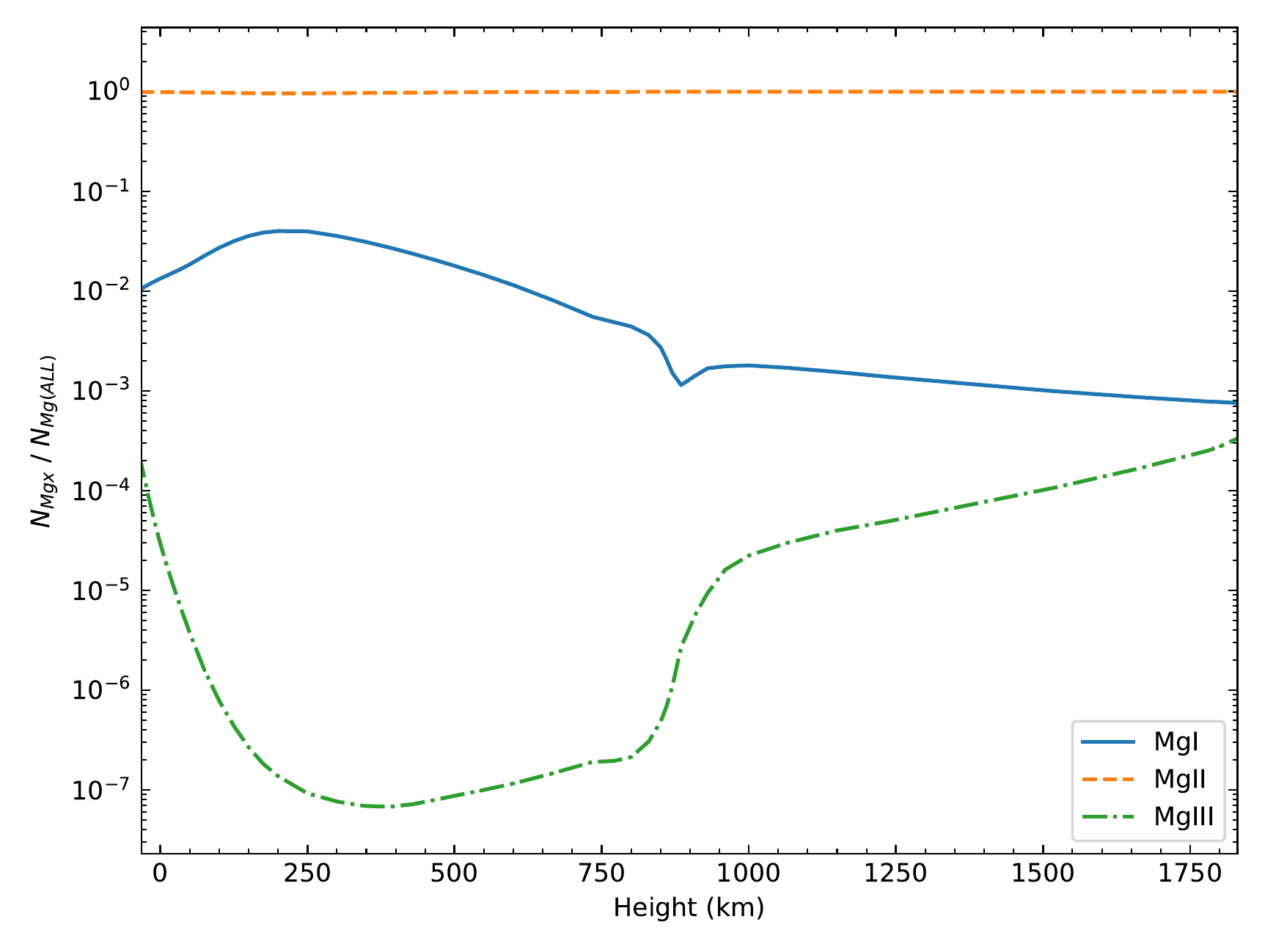}}
\caption{\MgI (blue line), \MgII (orange dashed line) and \MgIII (green dash-dot line) densities relative to the total magnesium density for atmospheric model 1401 as a function of height.}
\label{fig:mgtot}
\end{figure}

\section{The set of new atomic models} \label{sec:new_model}

The aim of the present work is to improve the \MgI calculated features in the solar spectra using the mentioned atmospheric model 1401. To this end, we built a set of variants of the model 1401 with a different \MgI model, but maintaining the same temperature vs. height atmospheric structure. We also modified the \MgII atomic model, but the rest of the atomic and molecular species were maintained in every variant. In this way, we were able to detect variations in the calculated spectra due to the imposed changes to the \MgI atomic model.

Table \ref{table:models} lists the models built and its characteristic modifications, which are described in detail in the following.

\subsection{Updating the atomic data (Model 1401a)} \label{subsec:atomic_updates}

New experiments and theoretical developments continue to provide more and new accurate atomic data. This information is continuously being updated in public databases. Consequently, atomic models must also be updated to obtain reliable spectra.

Model 1401a was built to identify the changes in the spectrum due to updated atomic data, maintaining the 26 energy levels structure for the \MgI atomic model, and the original 82 transition lines. The update was focused on the oscillator strengths (\textit{gf}), Einstein coefficients (\textit{A}), and broadening parameters of the spectral lines considered. \textit{gf} and \textit{A} were obtained from \textit{NIST} database version 5.7.1 \citep{NIST:2020}. Radiative, Stark, and van der Waals broadening parameters, initially missing in our database, were taken from \textit{VALD3}\footnote{\url{http://vald.astro.uu.se}} database (\cite{vald:2000}) and, when it was absent in \textit{VALD3}, from \textit{Kurucz} database \citep{kurucz:1995}\footnote{\url{http://kurucz.harvard.edu/linelists/gfall/}}.

\begin{figure}
\resizebox{\hsize}{!}{\includegraphics{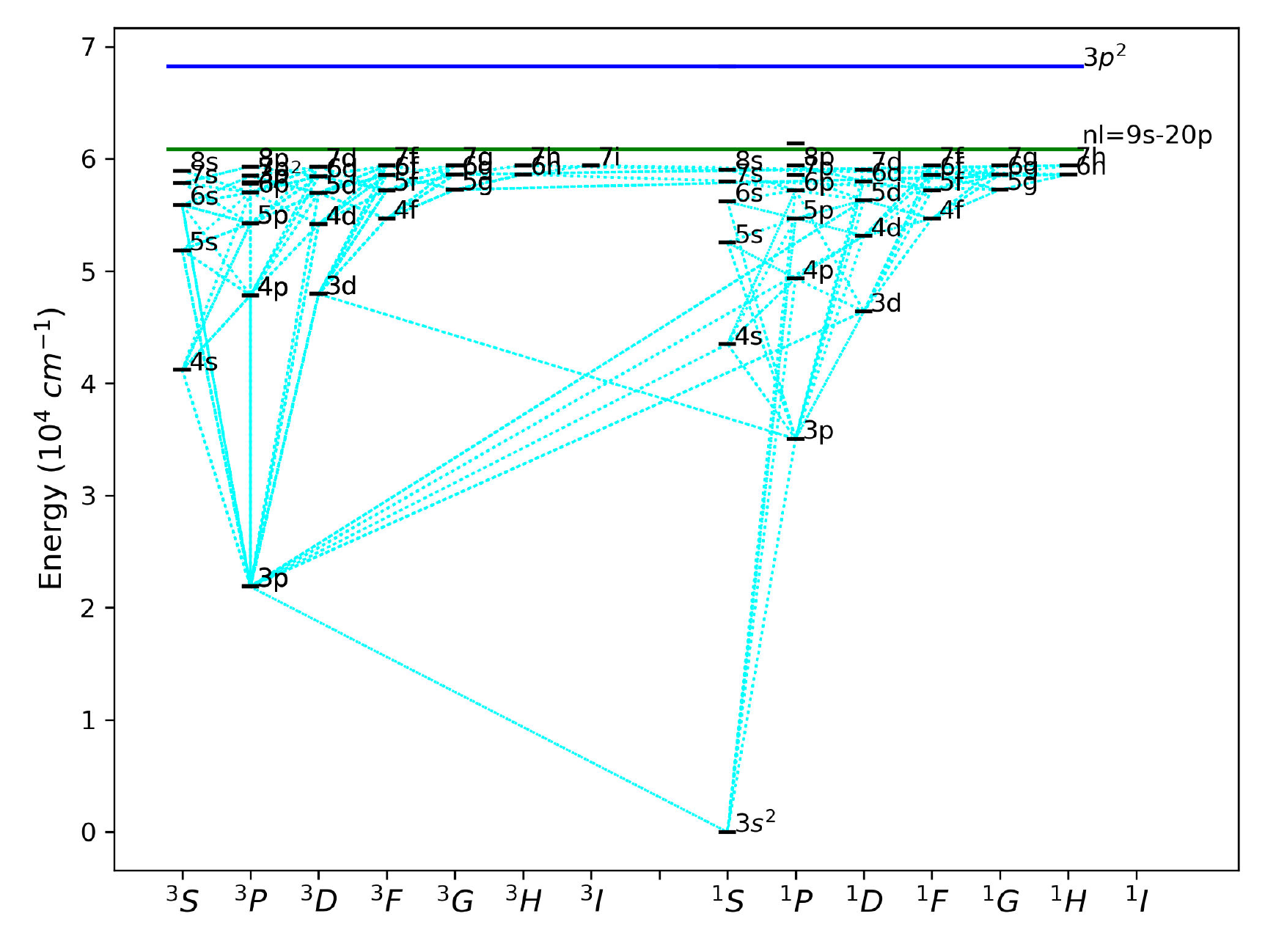}}
\caption{\MgI Grotrian diagram for the structure of 85-levels models. Fine structure is included up to level $7i\,^1I$ (level 54, 59\,430.517 $cm^{-1}$) (not distinguishable in the figure). First 54 energy levels are fully represented. Superlevels (between $nl=9s$ and $20p$) are shown with their average energy as a green-solid line, very close to the continuum (which occurs at 61\,671.05 $cm^{-1}$). Level $3p^2$ (68\,275 $cm^{-1}$) is represented as a blue-solid line. In cyan dashed lines are shown the 210 transitions included in the spectrum.}
\label{fig:grotriam}
\end{figure}

We updated 71 values of \textit{A} in total.
Figure \ref{fig:achanges} shows the distribution of changes in this parameter throughout the spectral range. The maximum change obtained was of 112\% in 6\,320.46 {\AA} $(4s\,^3S_1$--$6p\,^3P_{0,1,2})$, meanwhile the minimum was of 0.8\% in 2\,852.96 {\AA} $(3s^2\,^1S$--$3p\,^1P)$. It is important to note that changes in the \textit{A} parameter implied a re-computation of the collisional $\Upsilon_{ij}$ data obtained using SEA\&VRM formulas.
Regarding the broadening parameters, the Stark and van der Waals data was filled for 24 transitions.

\subsection{Atomic structure. Including more lines (models 1401b and 1401c)} \label{subsec:levels_lines}

The 26-energy levels structure in the initial \MgI atomic model (used in atmospheric model 1401 and 1401a) limits our capacity in reproducing strong lines in the NUV, and a major number of lines, higher than 17\,000 {\AA}, in the IR. 
In addition, the original highest energy level of 57\,262.76 $cm^{-1}$ $(3s5g\,^1G)$ was very far from the next ionization state \MgII, at 61\,671.05 $cm^{-1}$.
Figure \ref{fig:mgtot} shows the distribution of \MgI, \MgII and \MgIII relative to the total abundance of Mg at each height, throughout the atmosphere of the model 1401. 
\MgII is clearly the most abundant ionization state. As the majority of the population is concentrated in the \MgII ground-state, to obtain a realistic \MgI atomic model, it should be necessary to include Rydberg energy levels closely coupled with this ground-state. This effect was solved in our models 1401b and 1401c by significantly increasing the amount of energy levels, coupled to the lower states of \MgII by collisional ionization and recombination processes. 

In what follows, we detail the new energy structure for the \MgI and \MgII atomic models used in 1401b and 1401c atmospheric models, maintaining the updates made in the previous section to the first 26 levels and corresponding transitions.

\textbf{\MgI}. The energy structure was built to obtain an optimized atomic model capable of including new lines with minimal computational cost. The maximum energy level was increased to $3s20p\,^1P$ (61\,365.55 $cm^{-1}$) using values from the \textit{NIST} database, as in the \MgI models from \cite{osorio:2015} and \cite{alexeeva:2018}, although a previous version of the \textit{NIST} database was used in these works. Levels $3p^2\,^1S$ (68\,275 $cm^{-1}$) and $3p^2\,^3P_{0,1,2}$ (57\,812.77 $cm^{-1}$) were also included. We incorporated the fine-structure splitting for all levels up to $3s7i\,^3I$ (59\,430.517 $cm^{-1}$). Levels $3snl$ with $n\geq8$ and $l\geq2$ with the same main quantum number $n$ were merged into one state (e.g. $8d$ to $8h$, $9d$ to $9k$, and so on). For $n\geq9$, superlevels were also considered when $l=s,p$, to combine singlets with triplets in higher levels. 

The structure of our final \MgI model was formed by 85 levels and 129 sublevels. This allows one to obtain transitions involving Rydberg states, and also population exchange from the next ionization states (\MgII, \MgIII, etc). Figure \ref{fig:grotriam} shows, through a Grotrian diagram, the \MgI structure of levels and sublevels. Superlevels are represented by one state as a green-solid line, at average energy. Singlets and triplets are separated for better visualization.

\begin{figure*}
\centering
\includegraphics[width=\textwidth]{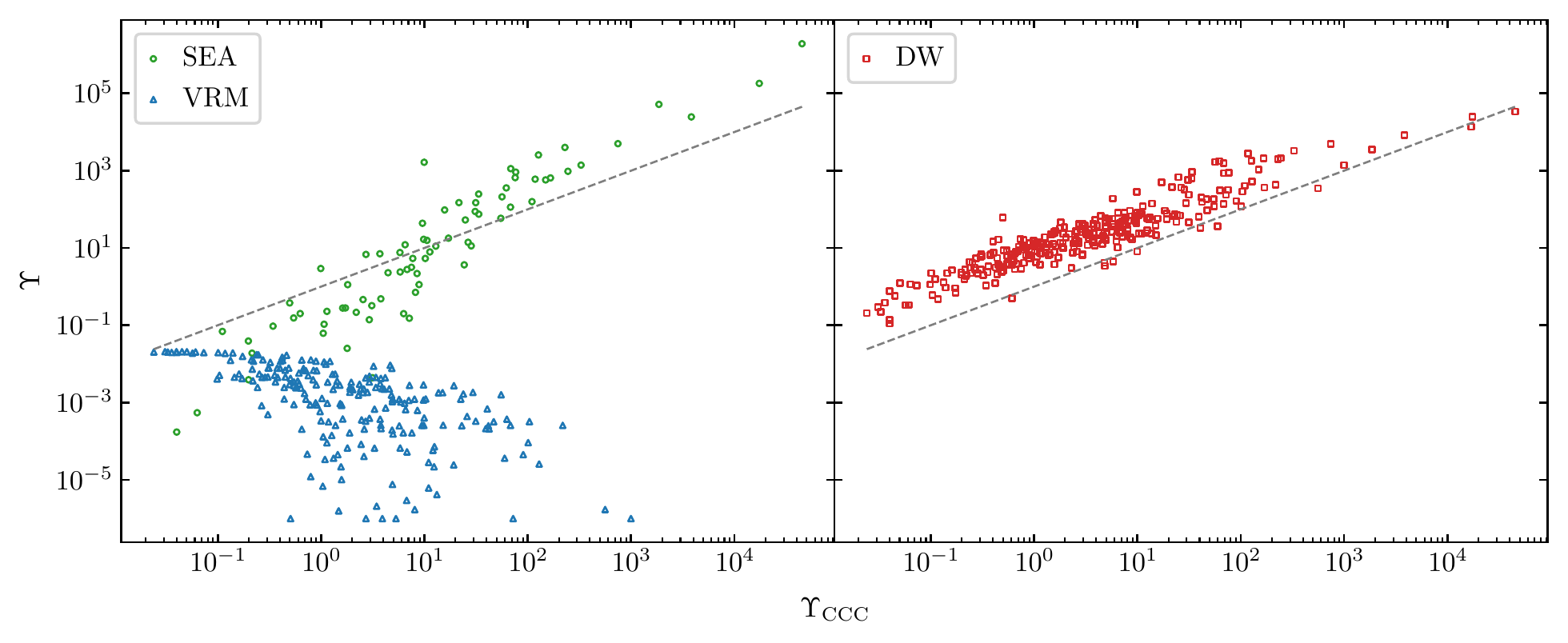}
\caption{Comparison of effective collision strengths $\Upsilon_{ij}$ at $T=6000$~K. \textit{Left}: CCC calculations vs. SEA (circles) and VRM (triangles) values. \textit{Right}: CCC data vs. DW (squares) results. Lines: CCC one-to-one relation (dashed).}
\label{fig:compECS_1-25}
\end{figure*}

To obtain the most relevant lines missing in the old atomic model, we performed a query on the \textit{NIST} database using the following criteria: $\log(gf)>-1$, maximum energy level 59\,649.15 $cm^{-1}$ ($3s9s\,^3S$), and a maximum wavelength (in vacuum) of 71\,500 {\AA}. As a result of this query, 127 lines were added to our database. This number is reduced to 58 if the distinction in quantum number \textit{J} is not considered. 

\MgI photoionization cross-sections currently available in the \textit{TOPbase} database were used in our new 85-level model, which includes transitions with $n\leq9$ and $l\leq4$. Consequently, the radiative recombination coefficients considered were also modified. Concerning the broadening parameters for the new lines, 33 transitions were completed with Radiative, Stark, and van der Waals parameters from \textit{VALD3} and \textit{Kurucz} databases. When the broadening parameters for the different possible processes for a certain transition were missing, the SRPM system solely considered the natural broadening parameter as $A_{ij}/4\pi$.

\textbf{\MgII}. For models 1401b and 1401c the \MgII atomic model was also updated. The number of energy levels was expanded from 14 ($6p\,^2P$, 105\,622.34 $cm^{-1}$) to 47 ($2p^6\,11g\,^2G$, 117\,639.51 $cm^{-1}$) using the \textit{NIST} database. The new structure included fine structure splitting without superlevels, resulting in a total of 90 sublevels. Consequently, we increased the number of transition lines from 52 to 781 (22 to 291 for term-term transitions), of which 405 lines were taken from the \textit{NIST} database and the remaining from the \textit{Kurucz} database. We were able to compute \MgII spectral lines in the enormous range of 850--6\,342\,878 {\AA}. The broadening parameters for the new lines were extracted from the \textit{Kurucz} database, when available. The electron impact parameters $\Upsilon_{ij}$  for the new collisional transitions were calculated employing the SEA\&VRM combination formulas. For ionizing collisions, the \cite{NRL:2005} equation was used. Finally, the photoionization (and the radiative recombination as the inverse process), data available in the \textit{TOPbase} database were used in our new 47-level model.

\subsection{Effective Collision Strengths \texorpdfstring{$(\Upsilon_{ij})$}{(ECS)} for \MgI in 1401b and 1401c models} \label{subsec:ecs}

\begin{table}
\centering
\begin{tabular}{|c|c|c|c|}
\hline
Parameters & DW   & SEA  & VRM \\
\hline
$\mu^*$    & 5.07 & 2.06 & $8.4\times 10^{-4}$ \\
\hline
$\sigma^*$ & 2.29 & 9.33 & 22.85 \\
\hline
\end{tabular}
\caption{Statistical parameters (geometric mean and standard deviation) of the ratio $\Upsilon/\Upsilon_{\mathrm{CCC}}$ for the semi-empirical models and perturbation calculations.}
\label{tab:statECS}
\end{table}

The benchmark excitation rate coefficients for \MgI are currently given by \cite{barklem:2017}. The complexity of computing the electron impact excitation of neutral atoms is widely known. For close-coupling methods, it requires the inclusion of a large number of states to account for the higher Rydberg states and the coupling to the continuum. For example, the current benchmark included 712 states, of which 680 were continuum pseudo-states. Despite a large number of terms, the atomic model considered enables one to correctly compute the excitation collision strengths only between the first 25 lower-lying levels.

\begin{table*}[t]
\centering
\begin{tabular}{ccccccc}
 \hline\hline
$\lambda_{vac}$ & Transition & Transition level numbers & $\log gf$ & $\log \Gamma_4/N_e$ & $\log \Gamma_6/N_H$ & Fig.\\ ({\AA}) & \, & \textit{L(SL)--U(SU)} & \, & $(rad\,s^{-1}\,cm^3)$ & $(rad\,s^{-1}\,cm^3)$ & \,\\
\hline
 2\,780.6 & $3p\,^3P_{0,1,2}$--$3p^2\,^3P_{0,1,2}$ & 2(1,2,3)--28(1,2,3) & 0.749 & -5.98 & -7.70 & \ref{fig:2780}\\ 
 4\,572.4 & $3s^2\,^1S_0$--$3p\,^3P_1$ & 1(1)--2(2) & -5.623 & -6.89 & -7.69 & \ref{fig:4572}\\ 
 6\,320.7 & $4s\,^3S_1$--$6p\,^3P_{0,1,2}$ & 4(1)--22(1,2,3) & -1.848 & -4.57 & -7.10 & \ref{fig:6320_10964}, \ref{fig:VIS+NIR}\\ 
 10\,964.4 & $4p\,^3P_{0,1,2}$--$5d\,^3D_{1,2,3}$ & 7(1,2,3)--21(1,2,3) & 0.091 & -3.40 & -7.10 & \ref{fig:6320_10964}, \ref{fig:VIS+NIR}\\ 
 33\,200.6 & $4d\,^3D_{1,2,3}$--$5f\,^3F_{2,3,4}$ & 13(1,2,3)--24(1,2,3) & 0.953 & -3.71 & -7.10 & \ref{fig:IR}a\\ 
 36\,807.6 & $5p\,^3P_{0,1,2}$--$5d\,^3D_{1,2,3}$ & 14(1,2,3)--21(1,2,3) & 0.74 & -3.40 & -7.10 & \ref{fig:IR}b\\ 
 38\,664.9 & $4f\,^1F_3$--$5g\,^1G_4$ & 15(1)--26(1) & 0.949 & -3.71 & -7.10 & \ref{fig:IR}c\\ 
 38\,669.1 & $4f\,^3F_{2,3,4}$--$5g\,^3G_{3,4}$ & 16(1,2,3)--27(1,2) & 1.199 & -3.71 & -7.10 & \ref{fig:IR}c\\ 
 38\,670.4 & $6s\,^3S_1$--$7p\,^3P_2$ & 18(1)--33(3) & -0.837 & -3.01 & -6.90 & \ref{fig:IR}c\\ 
 71\,092.0 & $5f\,^1F_3$--$6g\,^1G_4$ & 23(1)--37(1) & 0.866 & -3.01 & -7.00 & \ref{fig:IR}d\\ 
 71\,097.4 & $5f\,^3F_{2,3,4}$--$6g\,^3G_{3,4,5}$ & 24(1,2,3)--38(1,2,3) & 1.345 & -3.01 & -7.00 & \ref{fig:IR}d\\ \hline\hline
\end{tabular}
\caption{Term-term line transitions shown in this work. Wavelengths and oscillator strengths (\textit{gf}) were extracted from the \textit{NIST} database. Broadening parameters of the lines shown were extracted from \cite{kurucz:1995}. $\Gamma_4$ and $\Gamma_6$ are given at 5\,000 K.}
\label{table:lines}
\end{table*}

The spectral lines studied in this work involve transitions in the NUV and MIR, which correspond to levels higher than the $3s6s\,^1P$ term. However, there are no close-coupling calculations currently available for these levels. Since the transitions between higher levels are difficult to calculate within the close-coupling framework, we resorted to a perturbation approximation. In particular, a multi-configuration Breit-Pauli distorted-wave (DW) method was employed to compute the effective collision strength for a certain number of levels. It is well documented that perturbation techniques tend to overestimate $\Upsilon_{ij}$, and this effect is even greater in neutral targets. Nonetheless, we consider that these quantum calculations could represent a significant improvement over the semi-empirical ones. The present DW results were obtained employing the \textsc{autostructure} code\footnote{\url{http://amdpp.phys.strath.ac.uk/autos}} using non-relativistic radial wave functions in intermediate coupling. Theoretical details of the method can be found in \cite{badnell:2011} and references therein. For models 1401b and 1401c, the electronic structure considered includes electronic configurations up to $3s10f$ with 90 terms, while the first 54 terms were adjusted to the energy levels given by the \textit{NIST} database. The electronic configuration selected allowed us to compute the $\Upsilon_{ij}$ for the first 54 terms. However, the computation of transitions involving higher-lying terms could be considered in a future work since it would only signify considering a larger electronic structure and, hence, more time-consuming calculations.

The present improvements in the electron impact data are summarized in Table \ref{table:models}. The $\Upsilon_{ij}$ data considered initially for the models 1401 and 1401a were obtained using SEA\&VRM semi-empirical models for allowed and forbidden transitions, respectively. Two significant improvements were made to the new model regarding the $\Upsilon_{ij}$ data. First, the latest convergent close-coupling (CCC) data by \cite{barklem:2017} were included in the atomic model for the first 25 lower-lying terms, while the remaining values of SEA\&VRM for higher levels were maintained. This combination is included in model 1401b. Then, in model 1401c, we further included perturbation distorted method calculations for transitions between levels 26 ($3s5g\,^1G$, 57\,262.76 $cm^{-1}$) and 54 ($3s7i\,^1I$, 59\,430.517 $cm^{-1}$) while the remaining values of SEA\&VRM for higher levels and superlevels (55 | 59\,649.15 $cm^{-1}$ to 85 | 68\,275 $cm^{-1}$) were maintained.

\begin{figure}[h!]
\centering
\resizebox{\hsize}{!}{\includegraphics{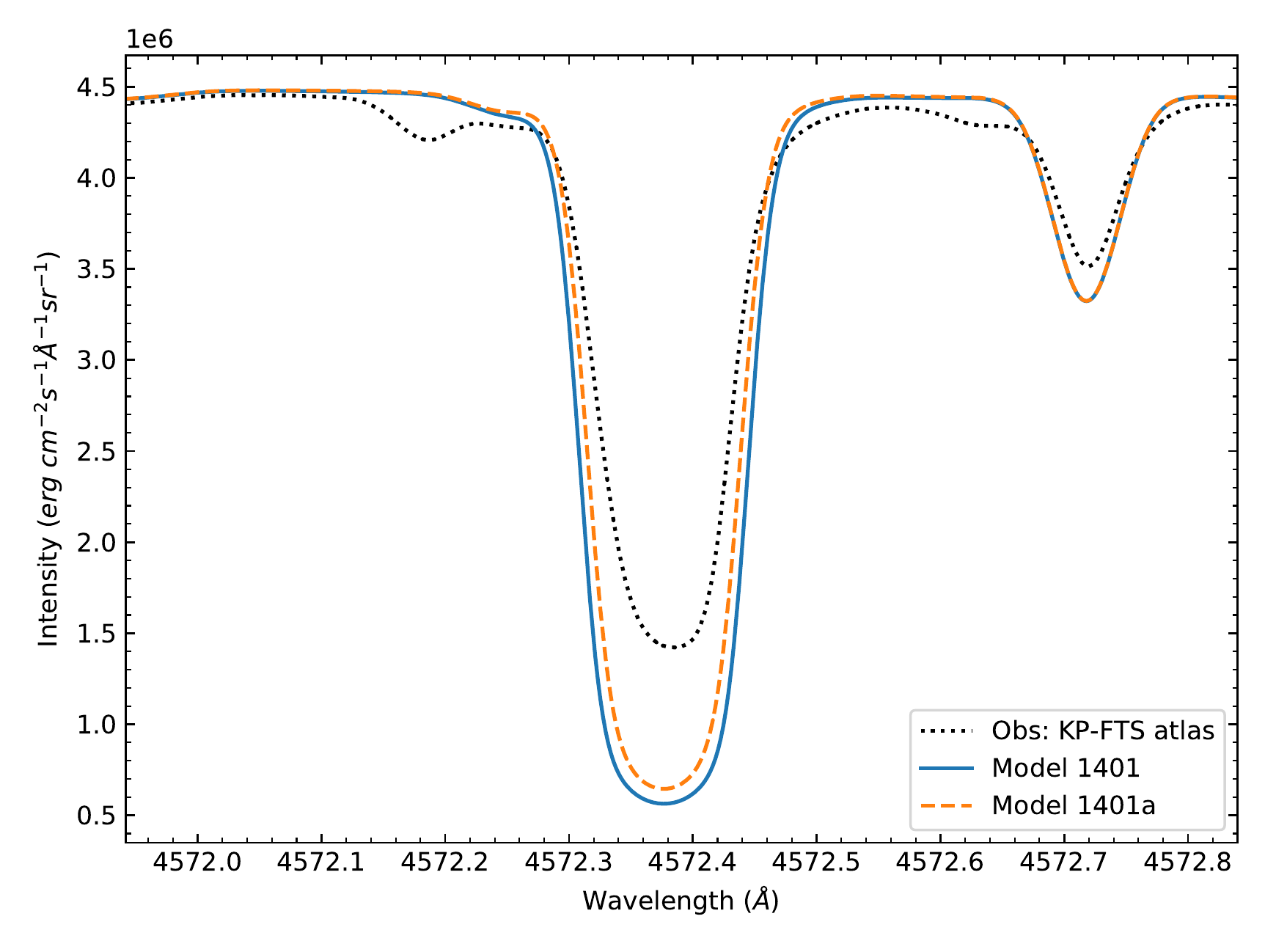}}
\caption{Models 1401 (blue line) and 1401a (orange dashed line) compared to Kitt Peak Solar Atlas (KP-FTS) \citep{acefts:2010} intensity observations (black dotted line). Transition 4\,572.4 {\AA} shows a small improvement due to an update in the oscillator strength parameter, whose value decreased by 41\%.}
\label{fig:4572}
\end{figure}

\begin{figure*}
\centering
\includegraphics[width=.5\textwidth]{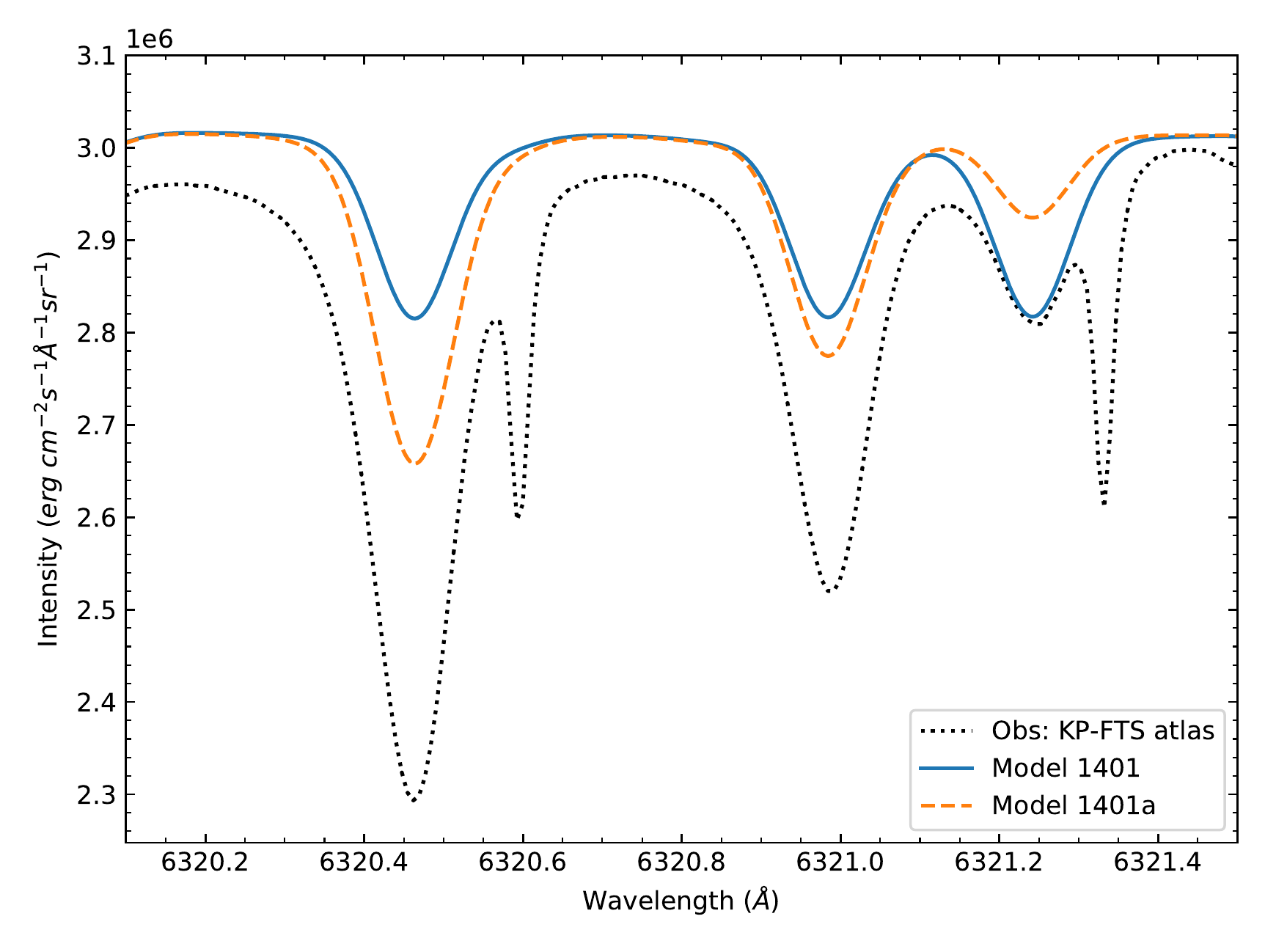}
\includegraphics[width=.5\textwidth]{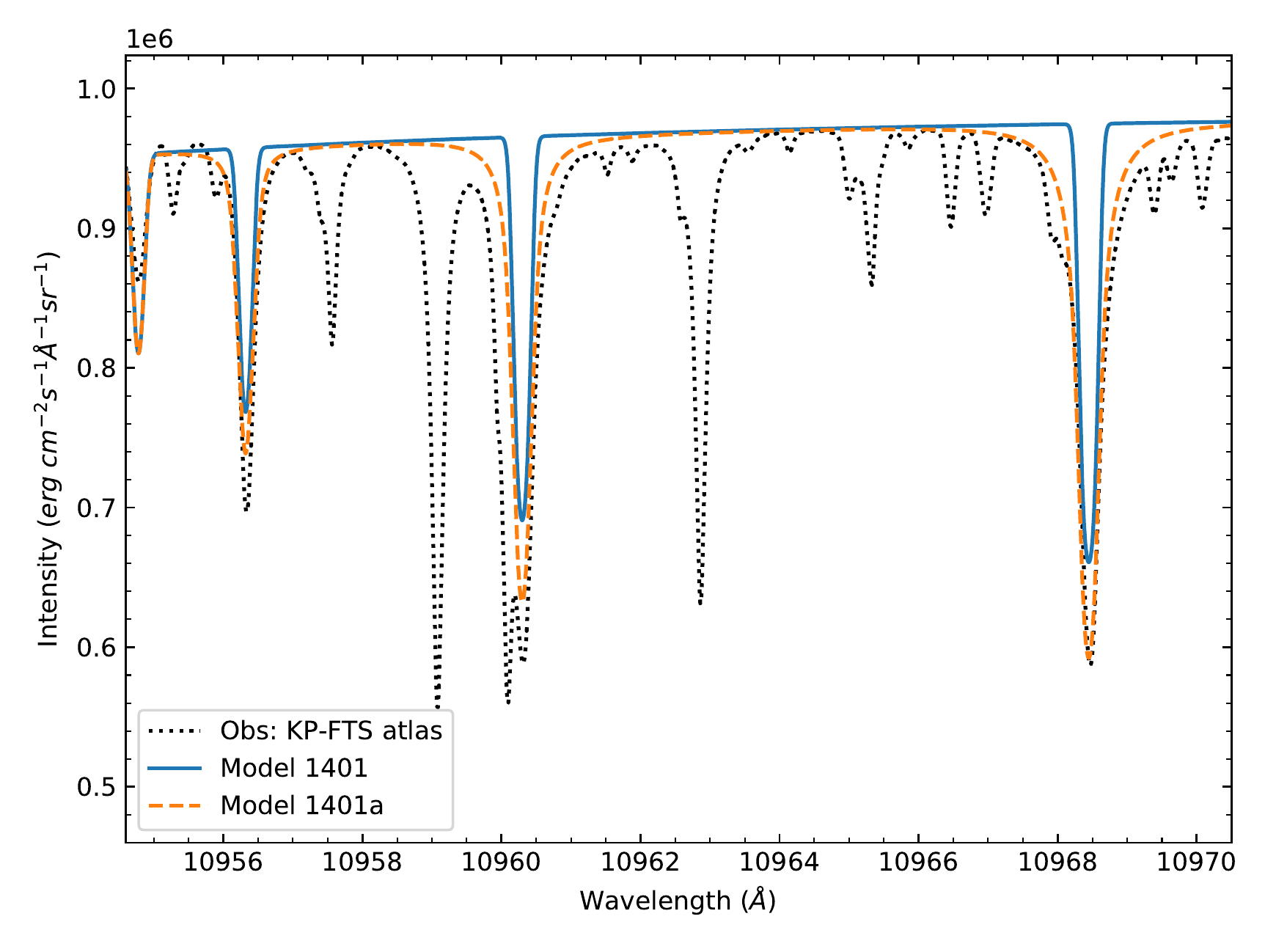}
\caption{Models 1401 (blue line) and 1401a (orange dashed line) compared to Kitt Peak Solar Atlas (KP-FTS) \citep{acefts:2010} intensity observations (black dashed line). Triplet transition around 6320 {\AA} (\textit{left panel}) shows improvements due to the update in the oscillator strength parameters, and the 10\,964.38 {\AA} profile (\textit{right panel}) shows improvements due to changes in the broadening data (not included in model 1401).}
\label{fig:6320_10964}
\end{figure*}

A qualitative comparison of the effective collision strength data sets considered in this work is presented in Fig. \ref{fig:compECS_1-25}. The figure shows the comparison of the $\Upsilon_{ij}$ values at $T=6000$~K given by the semi-empirical models considered (left panel) and the perturbation distorted wave calculations (right panel) with respect to the benchmark data (CCC) for the first 25 levels. The data sets from the semi-empirical model by SEA give a poor description for small and large transitions, while the VRM model is even worse, tending to underestimate the CCC values by several orders of magnitude with an opposite trend to the benchmark ones. The use of VRM formula has proven to be a generally bad approximation for the electron impact excitation cross-sections \citep{Sampson:92}. On the other hand, as expected, the perturbation calculations are systematically larger than the CCC results by a factor of 5. The ratio $\Upsilon/\Upsilon_{\mathrm{CCC}}$ for the perturbation method present a log-normal distribution, and the corresponding statistical parameters for all the approximations are shown in Table \ref{tab:statECS}. Inspecting the DW and the semi-empirical values from model 1401c, we found that the results employed follow a similar statistical behaviour.

\section{Observations} \label{sec:observations}
The \textit{SRPM} system produces a synthetic spectrum with an infinite spectral resolution that needs to be reduced to match the observed resolution data. To this end, the synthetic spectrum was convolved with a Gaussian function using a resolving power that depends on the resolution of observations used for comparison. In order to fairly judge the quality of the model and to take advantage of the information provided by the synthetic spectra, observations of mid to high resolution are required. The best solar observations we could obtain for each spectral range are described below.

\textit{Near-UV (NUV) range}. Solar observations spanning a wide range in the NUV, not only in specific spectral lines like the ones observed by IRIS spacecraft, are not easy to find. We used irradiance data from 200 to 310 nm from \cite{hall:1991} (hereafter H\&A). These observations were obtained from a stratospheric balloon that flew $\sim$40 km in the stratosphere. Although these data are difficult to calibrate, because the UV Fraunhofer spectrum is extremely rich in absorption lines, H\&A managed to combine and extrapolate it to zero optical depth to provide a reference spectrum with a spectral resolution of 0.01 nm.

\textit{Visible \& Near-Infrared (NIR) ranges}. In the range of 3290 to 12\,510 {\AA}, the selected observations were the Fourier-Transform-Spectra Solar Atlas (hereafter KP-FTS), obtained by Brault and co-workers at Kitt Peak Observatory \citep{neckel:1999}, as described by \cite{neckel:1984a,neckel:1984b} and \cite{neckel:1994}. According to \cite{neckel:1999}, the observations have no zero-point corrections to any of the FTS intensities, consequently, the absolute calibration for this disk-center $(\mu=1)$ spectrum is not reliable enough. For this reason, the calculated spectra were scaled (by adding a constant) to match the observed near continuum of the specific line profile under study. 

\textit{Mid-Infrared (MIR) range}. The data from ACE-FTS Solar Atlas\footnote{\url{http://www.ace.uwaterloo.ca/solaratlas.php}} \citep{acefts:2010} were selected. The ACE-FTS is a space-borne Fourier transform spectrometer onboard the SCISAT-1 spacecraft. This Solar Atlas is a solar transmission spectrum recorded at altitudes above 160 km that covers the range of 22\,500--130\,000 {\AA}, and contains no atmospheric contribution. In the cited work, the authors mention that the spectral range was built from an average of 224\,782 individual solar spectra, with a final 0.02 cm$^{-1}$ resolution.

\begin{figure*}
\centering
\includegraphics[width=.5\textwidth]{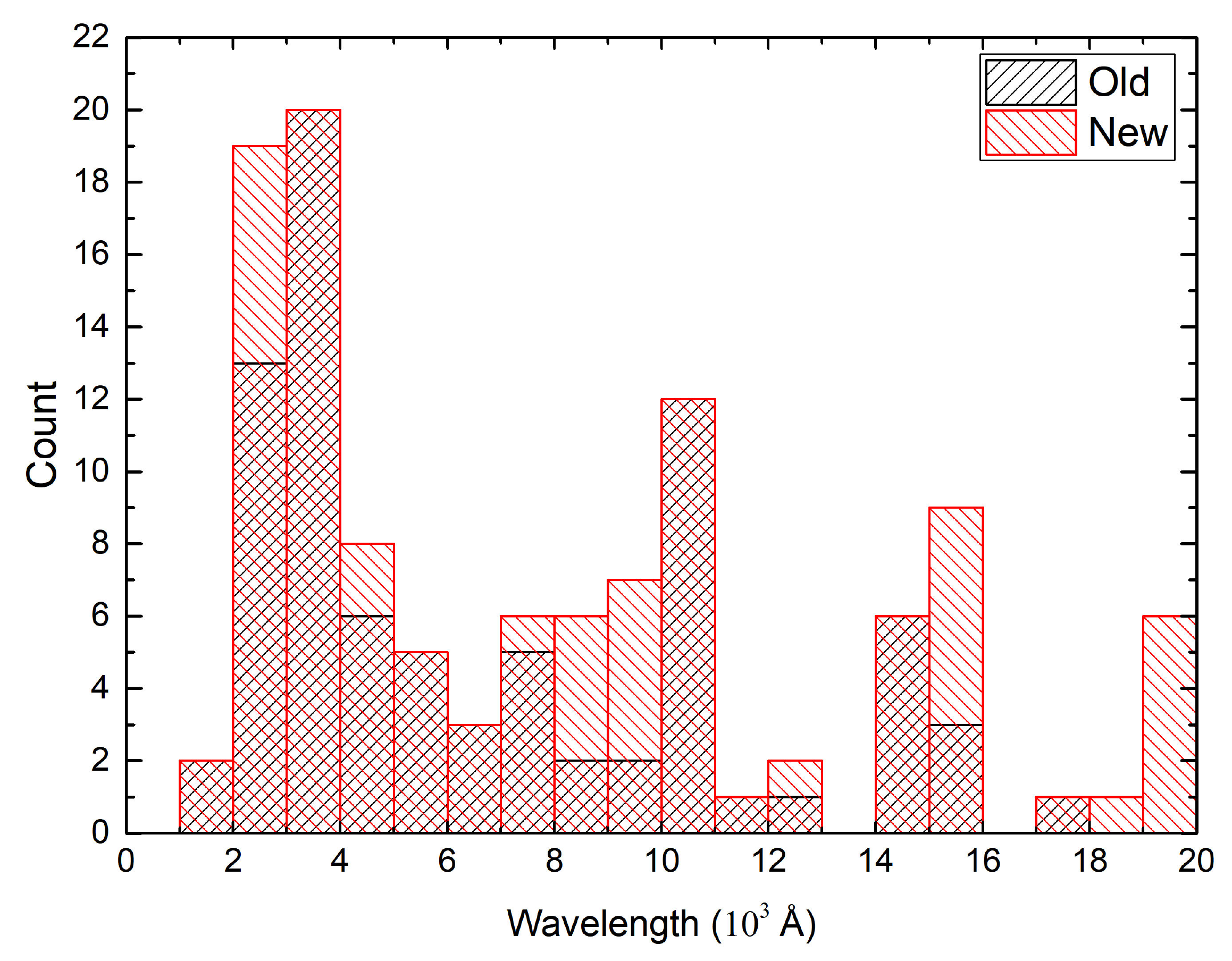}
\includegraphics[width=.5\textwidth]{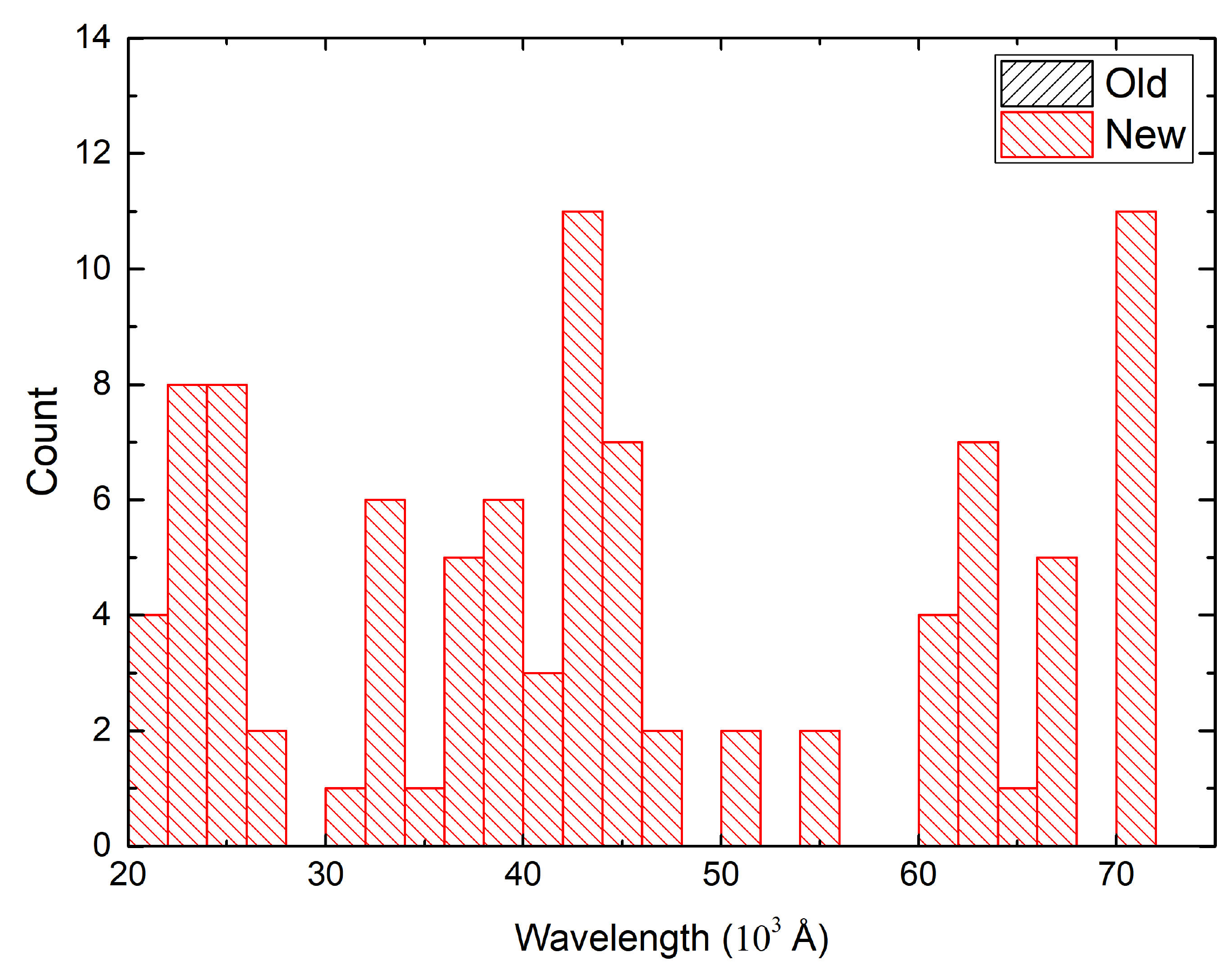}
\caption{Number of transitions included in the models. 127 new lines were added. \textit{Left}: NUV to NIR lines distribution for the 26-levels (black) and 85-levels (red) \MgI atomic models. \textit{Right}: Distribution of new lines in the IR range. There is no lines beyond 18\,000 {\AA} in models of 26 energy levels.}
\label{fig:lines_distrib}
\end{figure*}

\section{Results and Discussion} \label{sec:results}

\begin{figure}[h!]
\centering
\resizebox{\hsize}{!}{\includegraphics{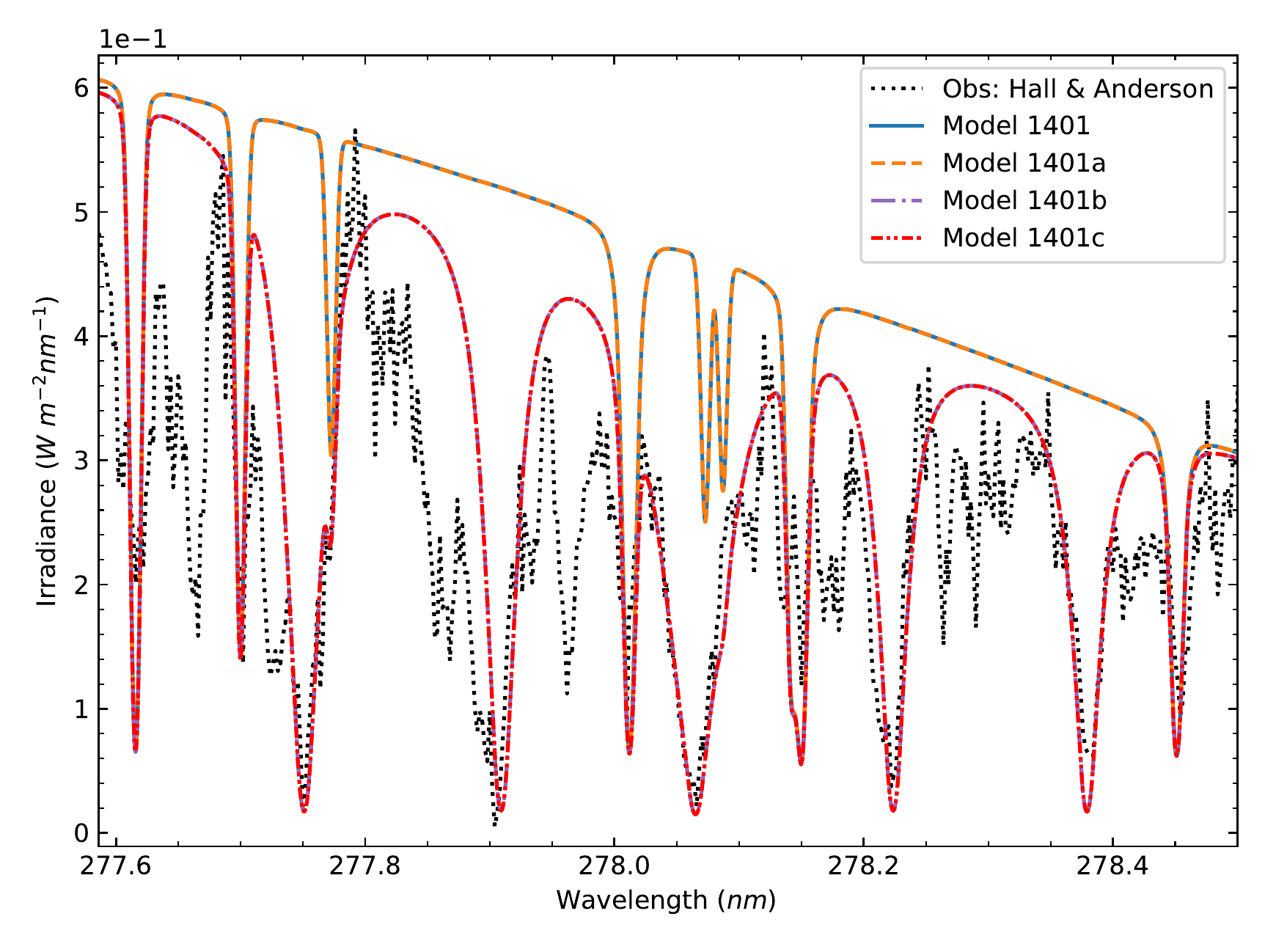}}
\caption{NUV spectral lines. Models 1401 (blue line), 1401a (orange dashed line), 1401b (purple dash-dot line) and 1401c (red dash-dot-dot line) vs. (H\&A) solar irradiance observations (black dotted line). New transitions 2\,780.65 {\AA} are included in 85-level models 1401b and 1401c due to the incorporation of the $3p^2\,^3P$ term.}
\label{fig:2780}
\end{figure}

\begin{figure*}
\centering
\includegraphics[width=.5\textwidth]{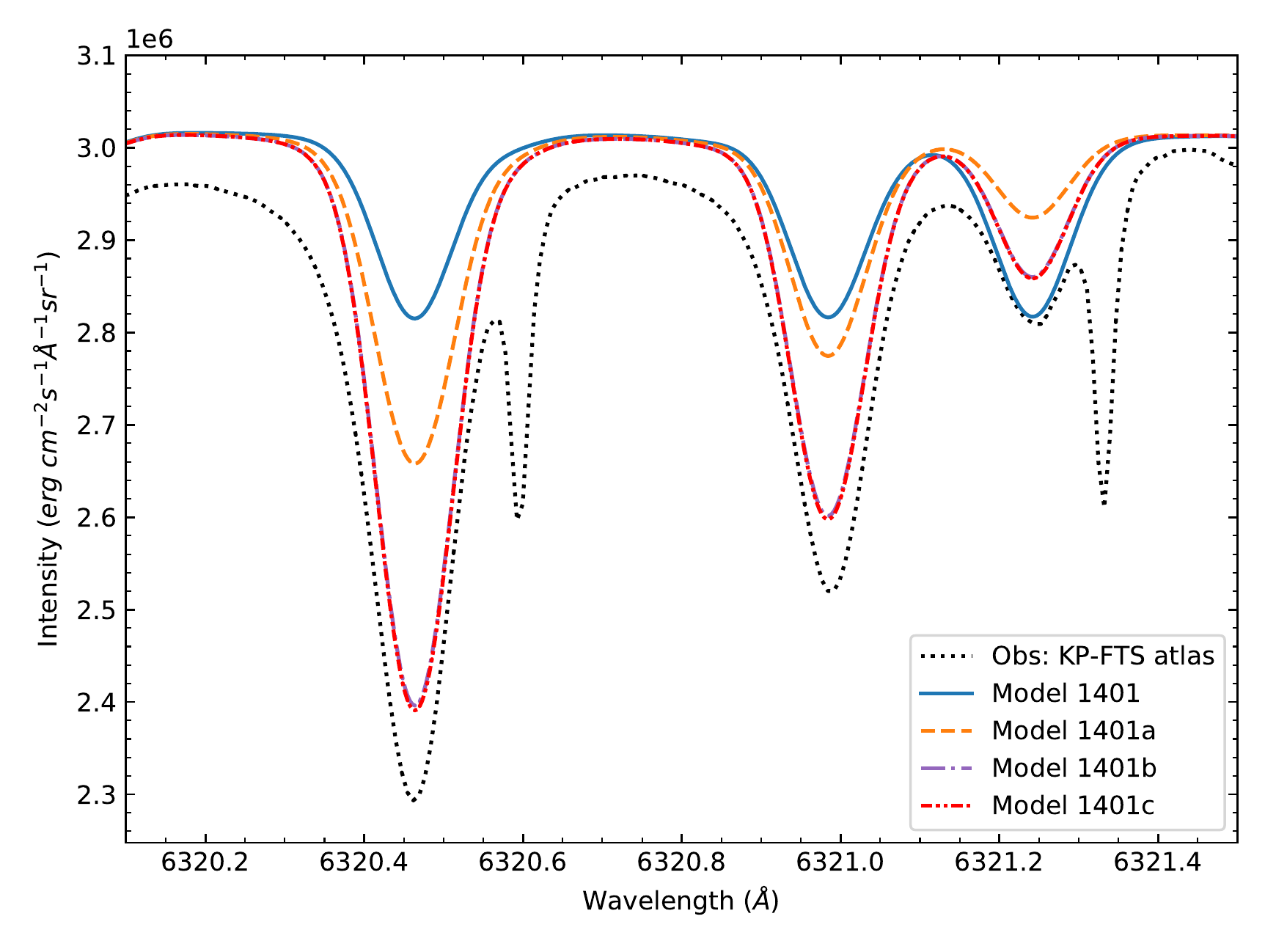}
\includegraphics[width=.5\textwidth]{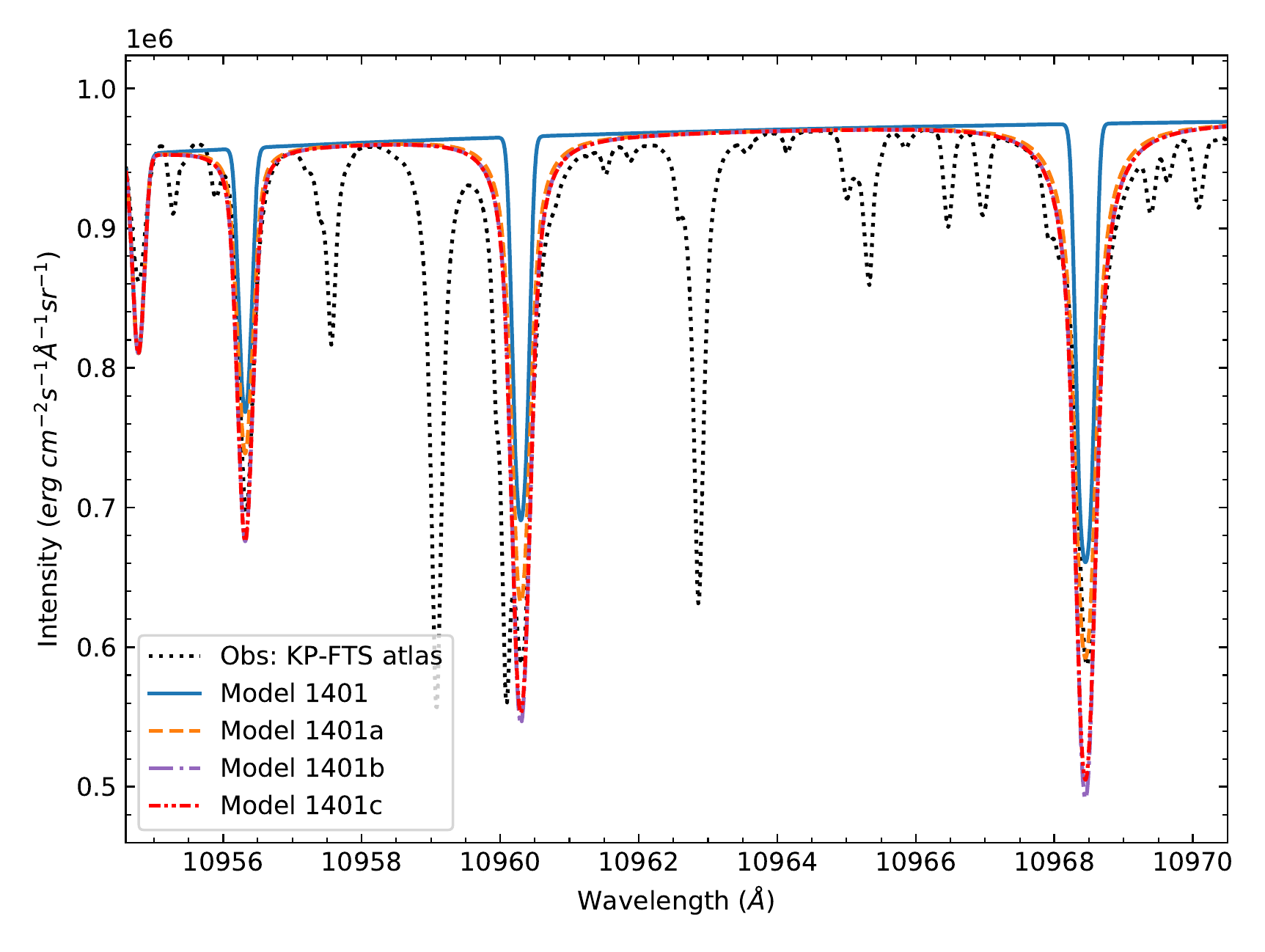}
\caption{VIS and NIR spectral lines. Models 1401 (blue line), 1401a (orange dashed line), 1401b (purple dash-dot line) and 1401c (red dash-dot-dot line) vs. observations (black dotted line). \textit{Left.} Visible range: transitions 6\,320.72 {\AA} compared to Kitt Peak Solar Atlas (KP-FTS) \citep{acefts:2010} intensity observations. Changes due to a higher number of levels  (models 1401b and 1401c) are clearly visible. \textit{Right.} NIR range: transitions 10\,964.38 {\AA} show improvements due to changes in the broadening data (not included in model 1401) and the increment in the number of energy levels (models 1401b and 1401c).}
\label{fig:VIS+NIR}
\end{figure*}

\begin{figure*}
\centering
\includegraphics[width=.5\textwidth]{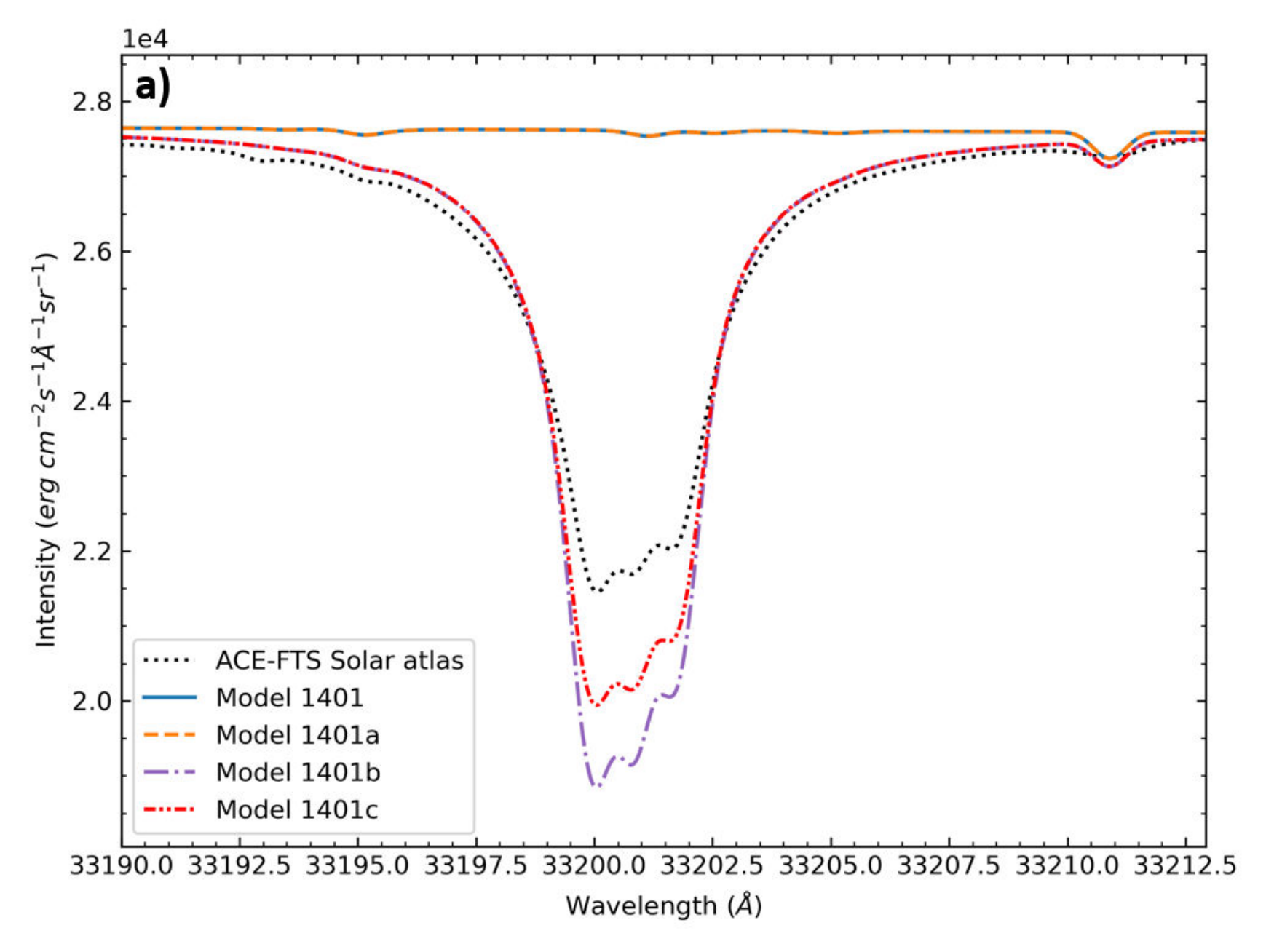}
\includegraphics[width=.5\textwidth]{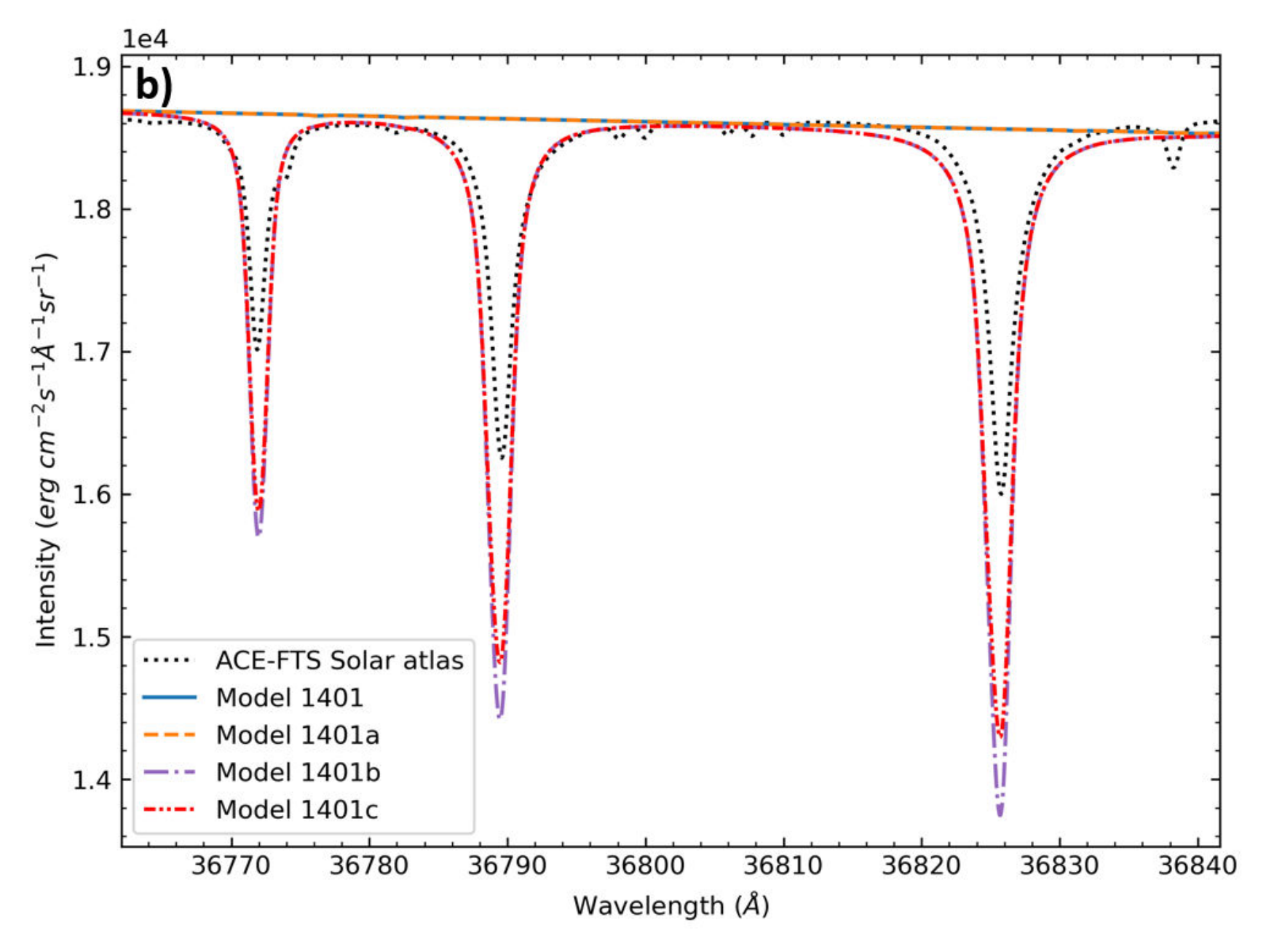}\vspace{0cm}
\includegraphics[width=.5\textwidth]{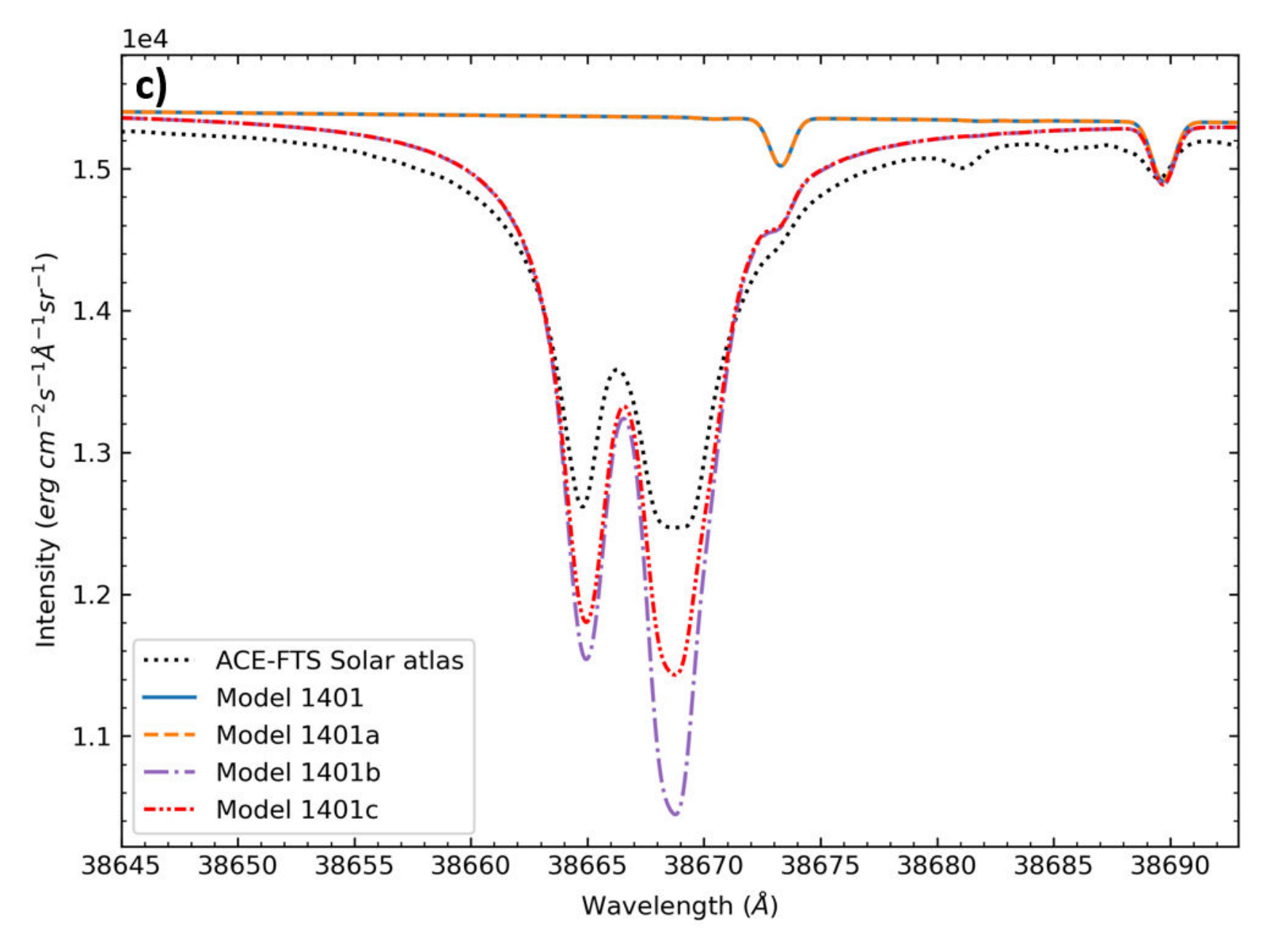}
\includegraphics[width=.5\textwidth]{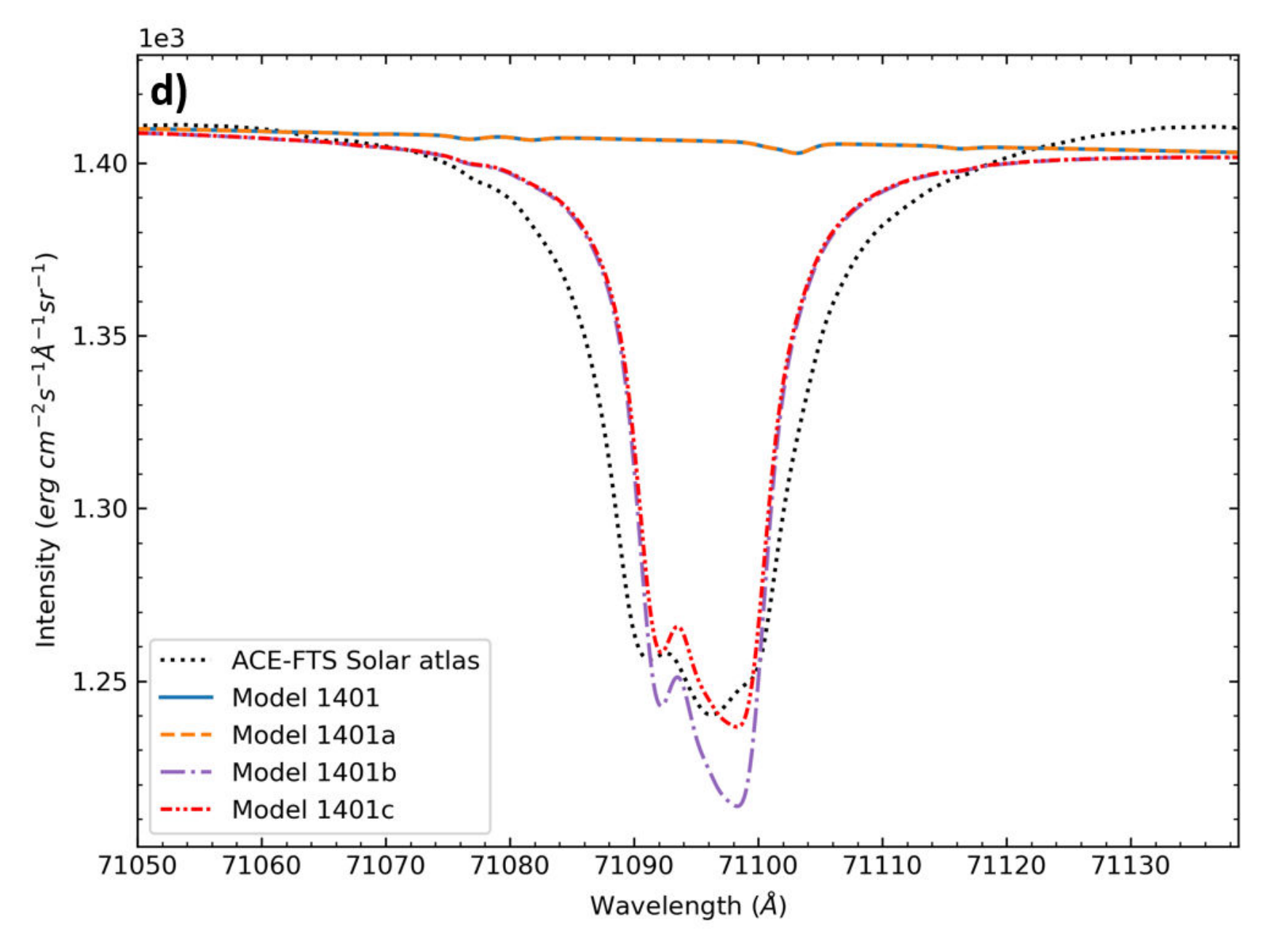}
\caption{IR computed spectral lines vs. ACE-FTS Solar Atlas observations (black dottedline). 26-levels models 1401 (blue line) and 1401a (orange dashed line) follow the continuum, 85-levels models 1401b (purple dash-dot line) and 1401c (red dash-dot-dot line) fit the observations. Term-term transitions: (\textit{a}) 33\,200.6 {\AA}, (\textit{b}) 36\,807.6 {\AA}, (\textit{c}) 38\,664.9, 38\,669.1 and 38\,670.4 {\AA}, (\textit{d}) 71\,092.0 and 71\,097.4 {\AA}. See Table \ref{table:lines} for line parameters. Although synthetic lines are deeper than observations, \MgI model 1401c with the new DW $\Upsilon_{ij}$ data, produces a better match throughout part of the IR range.}
\label{fig:IR}
\end{figure*}

After building the set of \MgI atomic models described in the previous sections, we run the atmospheric models 1401a, 1401b, and 1401c using the \textit{SRPM} system. We calculated the NLTE population for \MgI and all the other species present in the atmosphere. Once the converged populations have been reached, we computed the synthetic spectra for each case.

The aim of this study is to identify the changes in line formation related to the distinctive atomic model used in calculating the spectra. To this end, we have compared each new computed spectra with the synthetic spectra using the original model 1401 and solar observations, in several spectral ranges. In this section, we report the results of this comparison and analyze which of the \MgI atomic model tested improves the match with the solar observations.

The information about the \MgI lines we used to compare with observations in this section is presented in Table \ref{table:lines}. The last column in this table indicates the figure number where this transition is plotted. Several line profiles shown in the figures correspond to transitions between energy levels that were never before computed in full NLTE.

\begin{figure}[h!]
\centering
\resizebox{\hsize}{!}{\includegraphics{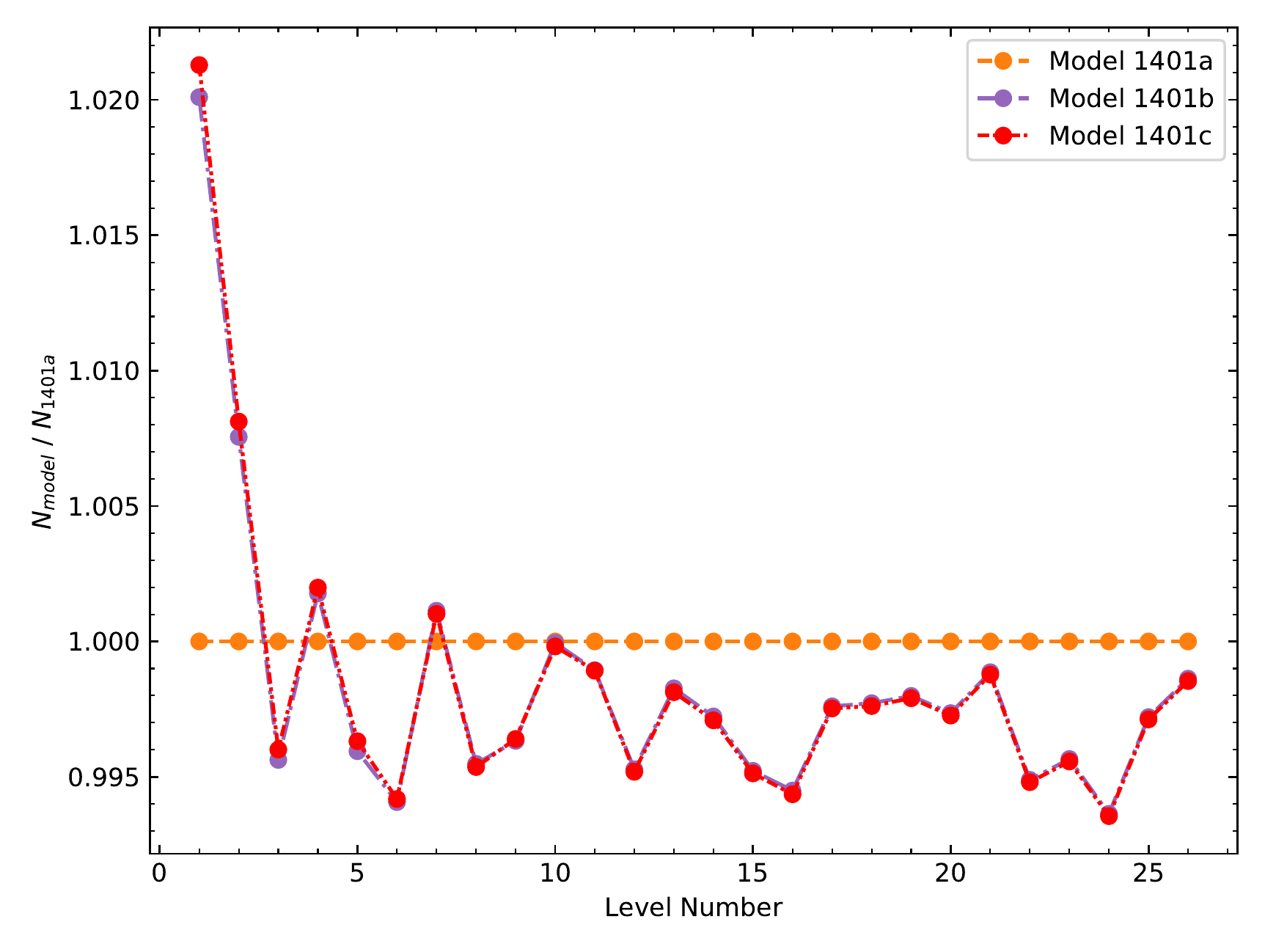}}
\caption{Level population ratio for model 1401b (purple connected dots) and 1401c (red connected dots) with model 1401 (orange connected dots). For each level, a sum over all heights in the atmosphere is performed.}
\label{fig:models_ratio_1401}
\end{figure}

\begin{figure}[h!]
\centering
\resizebox{\hsize}{!}{\includegraphics{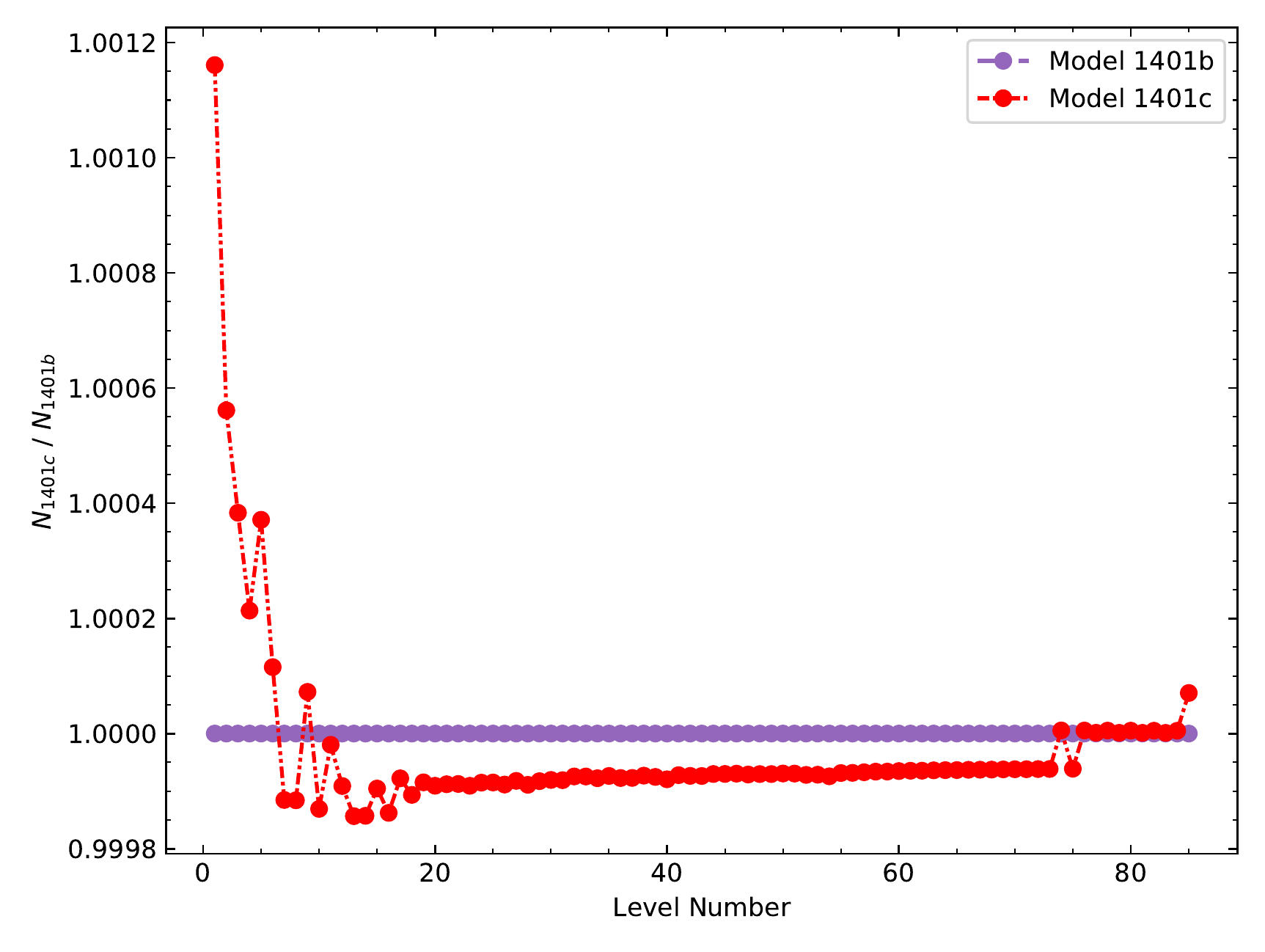}}
\caption{Level population ratio between models 1401c and 1401b. For each level, a sum over all heights in the atmosphere is performed.}
\label{fig:1401c_1401b}
\end{figure}

\subsection{Improvements owing to updated atomic data} \label{subsec:res_updates}

\begin{figure*}
\centering
\includegraphics[width=.5\textwidth]{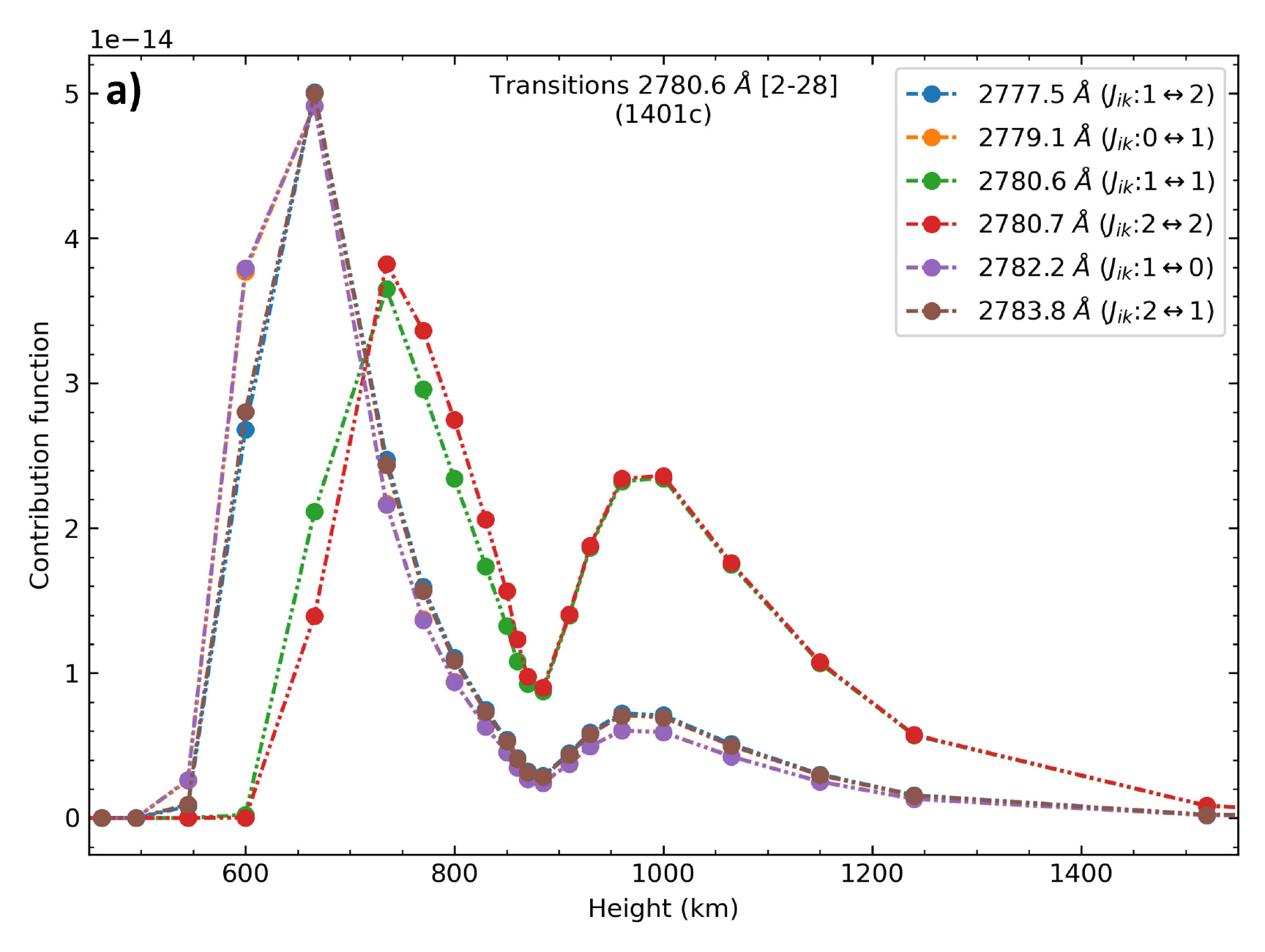}
\includegraphics[width=.5\textwidth]{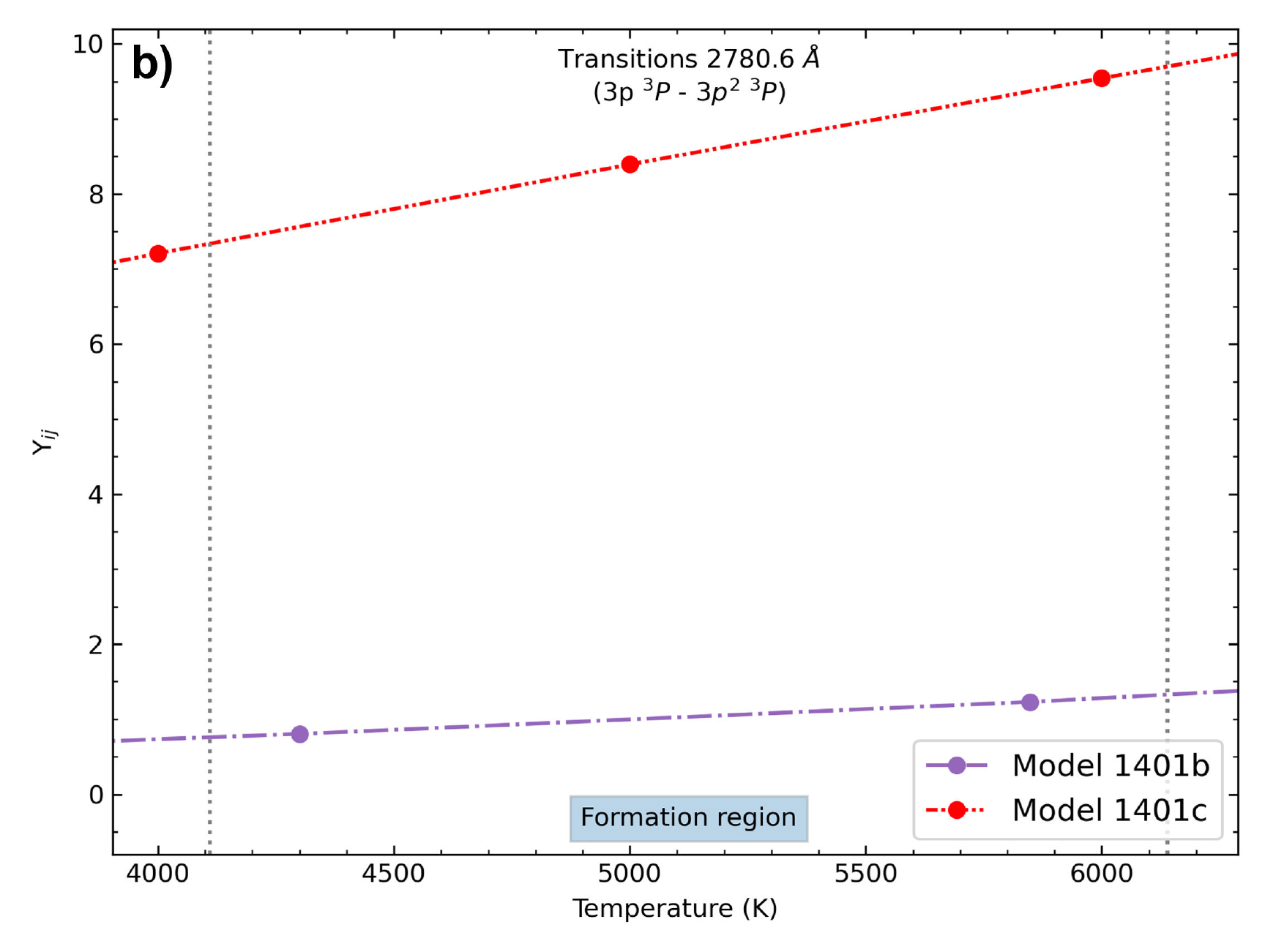}\vspace{0cm}
\includegraphics[width=.5\textwidth]{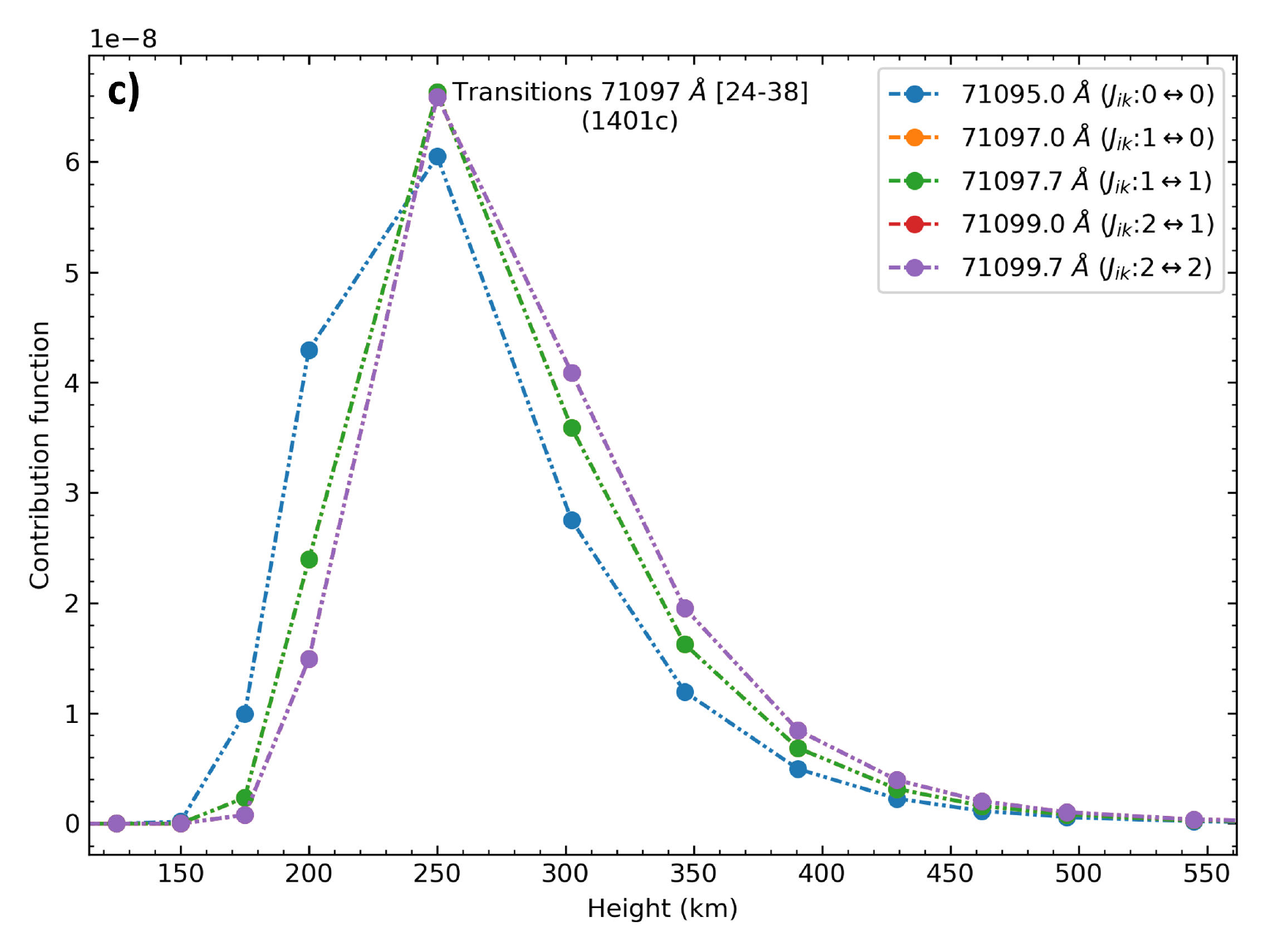}
\includegraphics[width=.5\textwidth]{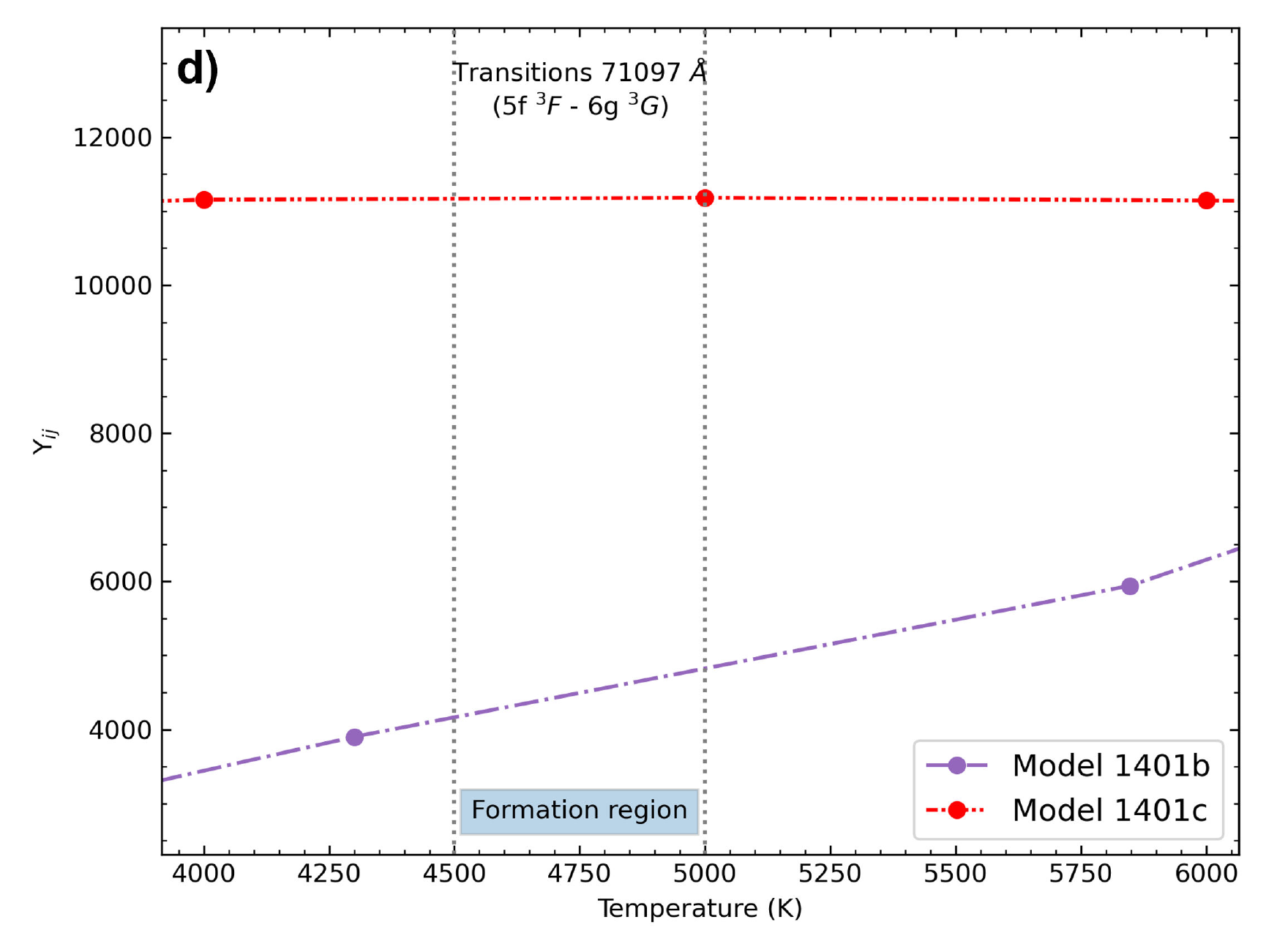}
\caption{\textit{Left panel (Figs. (a) and (c))}: 1401c model contribution functions through the atmospheric height, for lines involved in transitions 2\,780.6 {\AA} (a) and 71\,097 {\AA} (c). \textit{Right panel (Figs. (b) and (d))}: Effective Collision Strengths $(\Upsilon_{ij})$ data at the formation region temperatures that are used to compute the populations for 2780.6 {\AA} (b) and 71\,097 {\AA} (d) transitions, in models 1401b (purple dash-dot line) and 1401c (red dash-dot-dot line). $\Upsilon_{ij}$ data provided by the DW method are greater than the SEA\&VRM for all the transitions studied.
}
\label{fig:ECS_contrib}
\end{figure*}

We applied the updates described in Section \ref{subsec:atomic_updates} to model 1401a. The updated Einstein coefficients and broadening parameters produced noticeable changes, which improved the match with observations among the stellar spectra. Figures \ref{fig:4572} and \ref{fig:6320_10964} show examples of this improvement. 
The first figure presents a comparison for a line profile calculated with models 1401 and 1401a and the Kitt Peak solar Atlas. The computed line profile with model 1401a is slightly shallower than that calculated with model 1401, revealing the impact on the calculated line due to the updated oscillator strength, which decreased by 41\%. 

Both panels of Fig. \ref{fig:6320_10964} present triplet transitions, which are produced by a transition between different sublevels for the same upper and lower levels. The lines computed by model 1401a are deeper than those calculated using model 1401. Besides, each individual transition in the triplet calculated by 1401a has the correct change in depth following the observed lines, which is not present in the triplet obtained by 1401.
 
Calculated spectral lines in Fig. \ref{fig:6320_10964} (right panel) show the importance of considering a complete and accurate set of broadening parameters. In addition to the Radiative Damping, also the Stark and van der Waals coefficients were included in the atomic model of 1401a, producing wider wings that correctly match the KP-FTS Solar Atlas observations. 

Regarding \MgI density fraction, a small increase of 0.55\% was found in model 1401a relative to model 1401. This means that the updated made in the oscillator strength and Einstein coefficient have driven an overall population redistribution among \MgI, \MgII, and \MgIII fractions. The 99.987\% of this additional population of \MgI comes from \MgII and the rest from \MgIII. Notwithstanding, the relative distribution of these ions presents the same behavior shown in Fig. \ref{fig:mgtot}.

\begin{figure*}
\centering
\includegraphics[width=.5\textwidth]{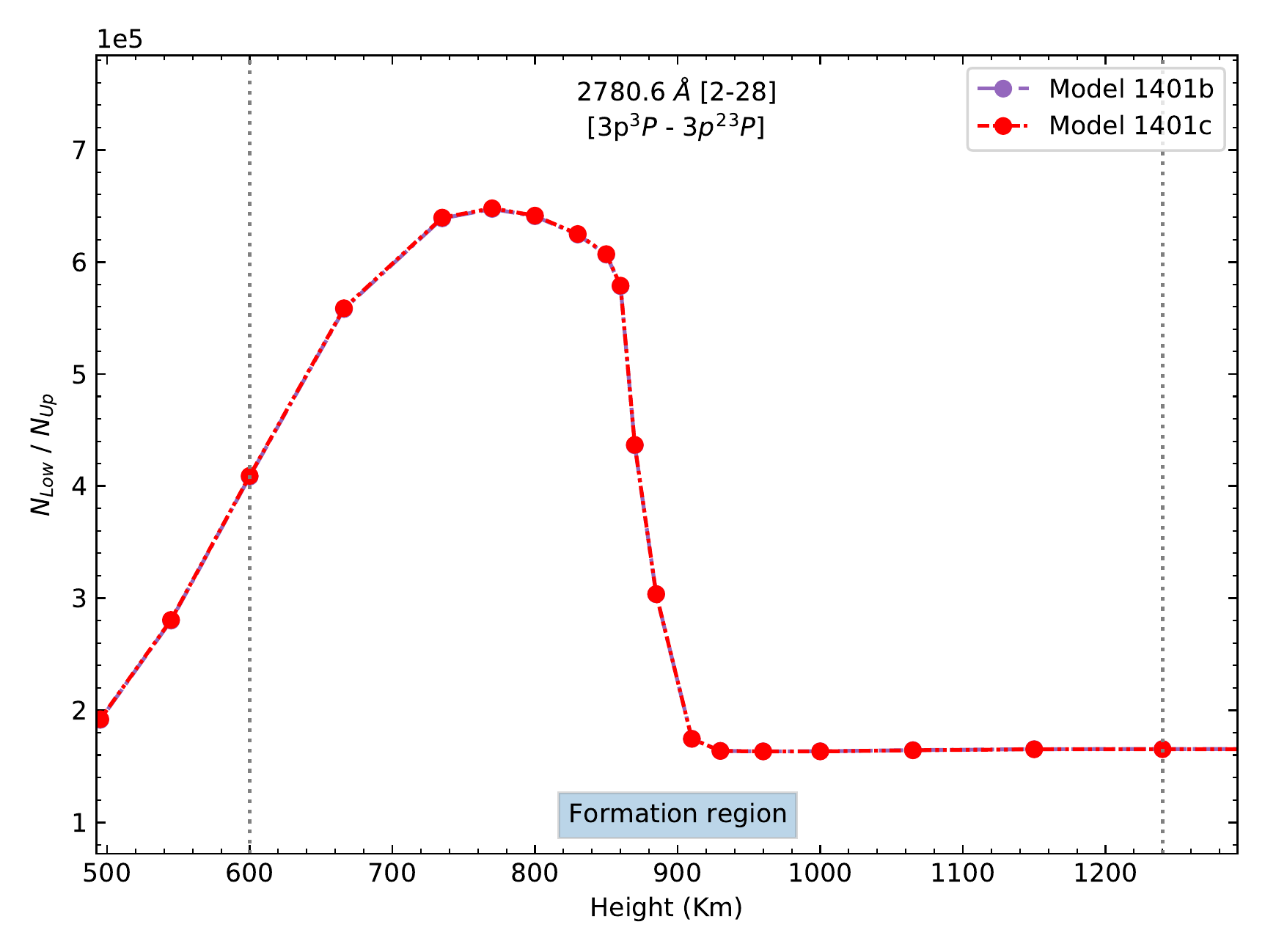}
\includegraphics[width=.5\textwidth]{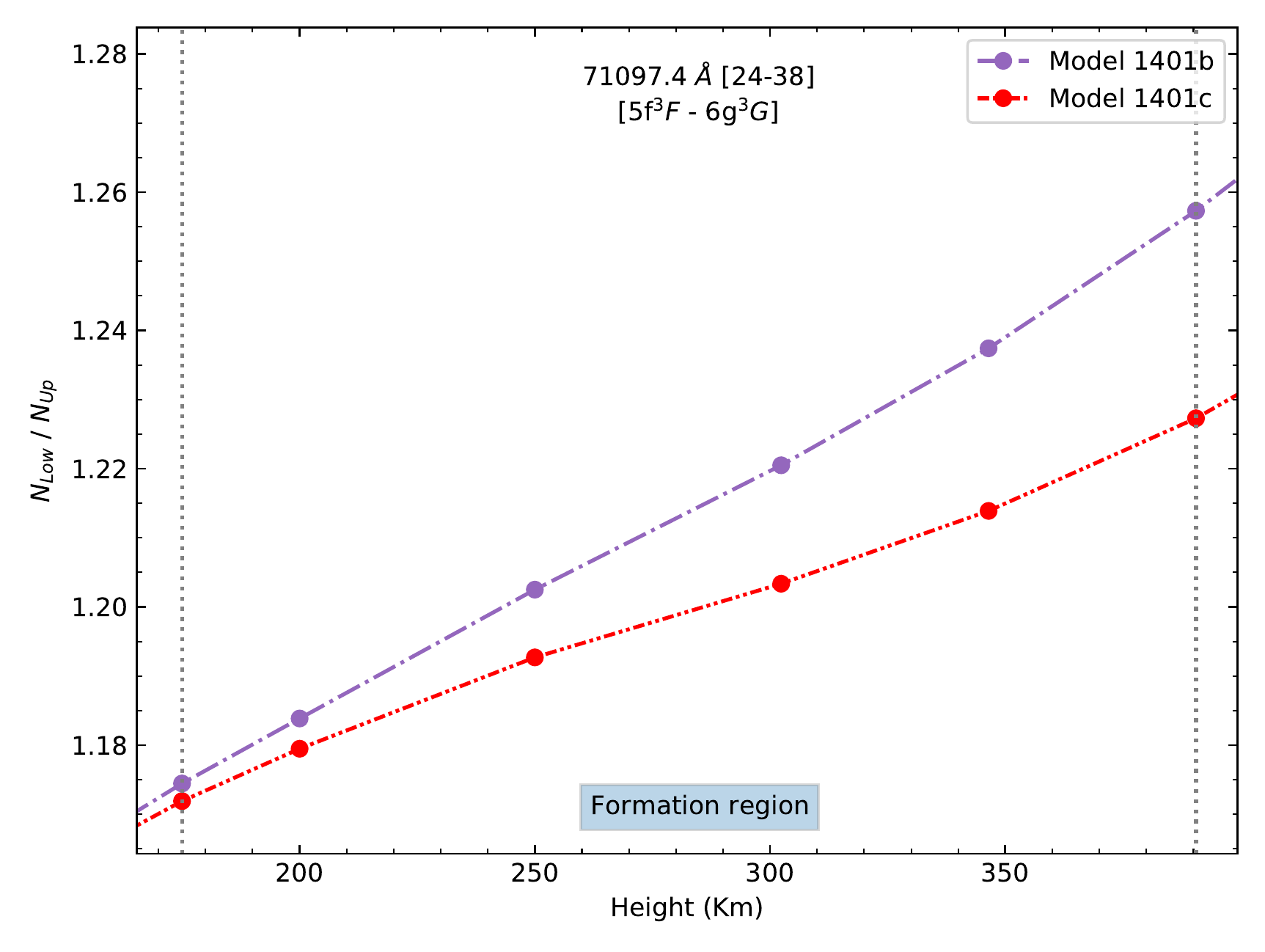}
\caption{Level population ratios $(N_{low}/N_{up})$ at formation heights for transitions 2\,780.6 {\AA} (left) and 71\,097 {\AA} (right), following the examples in Fig. \ref{fig:ECS_contrib}(a--b), and Fig. \ref{fig:ECS_contrib}(c--d), respectively. \textit{Left panel:} ratios are identical for models 1401b (purple dash-dot line) and 1401c (red dash-dot-dot line), producing the same line (shown in Fig. \ref{fig:2780}). \textit{Right panel:} as in almost every IR transition, levels population ratios are higher in model 1401b, producing deeper spectral lines (the line for this example is shown in Fig. \ref{fig:IR}d).}
\label{fig:levelsratio}
\end{figure*}

\begin{figure}
\centering
\resizebox{\hsize}{!}{\includegraphics{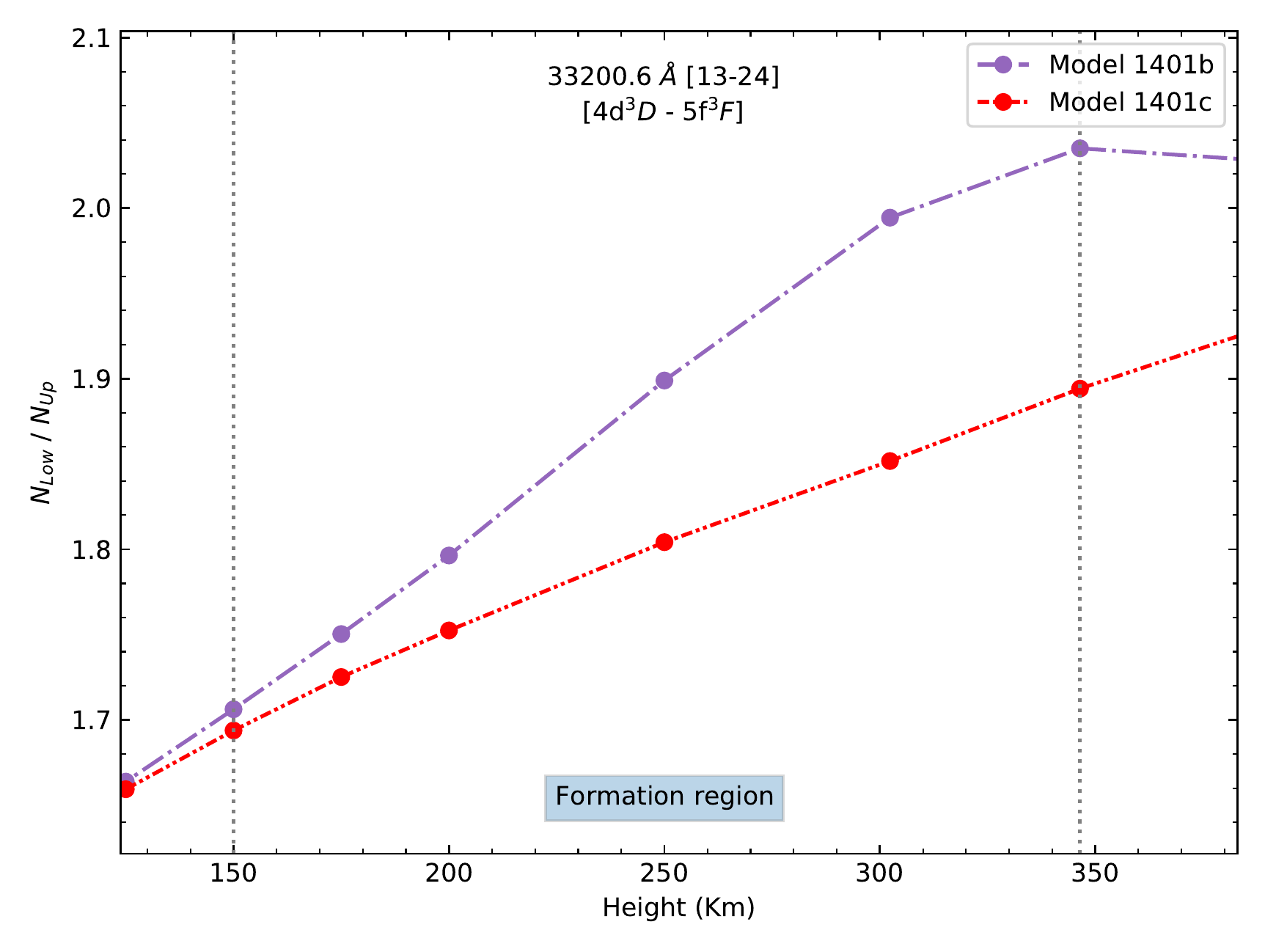}}
\caption{Levels population ratio for the 33\,200.6 {\AA} feature. Although both models used the same CCC $\Upsilon_{ij}$ data for the levels involved in the transition, the ratio changed due to the coupling with the rest of the levels in the atomic model.}
\label{fig:ratio_33200}
\end{figure}

\subsection{Including more levels and lines} \label{subsec:res_lineslevels}

Models 1401 and 1401a use an atomic \MgI model of 26 energy levels. This fact allowed only to calculate several transitions from the NUV to the NIR, but leaves aside other important transitions, specially in the IR. Examples of this limitation are shown in Figs. \ref{fig:2780} and \ref{fig:IR}, where the transitions plotted in each panel are produced between levels not included in these models.

To solve this restriction, and in addition to the previously mentioned updates, we built two new models. Both, 1401b and 1401c, include an increased number of energy levels (detailed in Section \ref{subsec:levels_lines}) to reproduce new spectral lines  across the studied range, from NUV to MIR (1747--71\,099 {\AA}). Although these models consider the same number of energy levels, there is an important difference between them. In the next section, we address this topic. 
The number of spectral lines added to our database in each spectral range, together with the total number of lines considered in our models 1401b and 1401c, are listed in Table \ref{tab:numLines}. Line distribution among the spectral range of interest is represented in Fig. \ref{fig:lines_distrib}.

\begin{table}
\centering
\begin{tabular}{|c|c|c|}
\hline
Spectral Range ({\AA}) & \# New Lines  & \# Total Lines \\
\hline
FUV (100--2000) & 0 & 2\\
\hline
NUV (2000--4000) & 6 & 39 \\
\hline
VIS (4000--7000) & 2 & 16 \\
\hline
NIR (7000--25\,000) & 39 & 73 \\
\hline 
MIR (25\,000--71\,100) & 80 & 80 \\
\hline
\end{tabular}
\caption{Number of new lines added to our models, and total lines included in each spectral range.}
\label{tab:numLines}
\end{table}

In Fig. \ref{fig:2780}, the incorporation of the $3p^2\,^3P$ term allowed 1401b and 1401c models to reproduce the NUV lines shown in this plot. The match between the two calculated spectra (which are overlapped in the plot) and the H\&A solar irradiance observations is remarkable. It is important to note that lines calculated using models with the same number of levels are overlapped, indicating that the change shown between calculated lines is triggered only by the number of levels considered.

The inclusion of more energy levels results in a general improvement of the calculated spectra. It permitted not only to reproduce new spectral features but, in addition, it improved the match in transitions between levels that have been already computed using a 26-levels atom. This is evident in Fig. \ref{fig:VIS+NIR}, where the lines calculated using 85-levels models have a remarkable improvement over the computed profiles with 26-levels models.

From the IR spectral lines added to our calculation, we selected four spectral features to illustrate the results. Figure \ref{fig:IR} shows the following transitions: (\textit{a}) 33\,200.6 {\AA}, (\textit{b}) 36\,807.6 {\AA}, (\textit{c}) 38\,664.9, 38\,669.1 and 38\,670.4 {\AA}, (\textit{d}) 71\,092.0 and 71\,097.4 {\AA}, compared with ACE-FTS Solar Atlas IR observations. The 1401 and 1401a models do not consider the line, and the calculated spectra follows the continuum, while the models 1401b and 1401c display an acceptable match with observations. Although both models 1401b and 1401c produce deeper spectral lines than observed, the atomic model 1401c generates shallower lines than 1401b, improving the match for all lines included in the studied range. The upgrade obtained using model 1401c in matching the observations is described in the next section.

As a consequence of adding more levels to the model, a redistribution of population among the ionized states of Mg occurred throughout the atmosphere. Each model follows the general behavior we have already presented in Fig. \ref{fig:mgtot}. However, we found an increase of 2.82\% and 2.93\% of \MgI for models 1401b and 1401c, respectively, relative to model 1401a. This additional population comes mainly from \MgII ($\sim$99.82\%) and the remaining from \MgIII ($\sim$0.18\%).
Compared with model 1401a, we obtained an increase of about 2.0\% and 2.1\% in the ground-state of \MgI in models 1401b and 1401c, respectively. This extra population comes mainly from \MgII and \MgIII ionized states and, to less extent, from \MgI higher levels. This last contribution is shown in Figure \ref{fig:models_ratio_1401}, where a decrease in the population for levels higher than the first two is evident in both models.

We attribute this result to a greater number of energy levels very close to the continuum, intentionally included to allow a flow of population from higher ionized to low-lying states, as it was suggested by \cite{osorio:2015} and \cite{carlsson:1992}. 
The redistribution of Mg population, as much throughout \MgI levels itself as among the ionized states fraction \MgI, \MgII, and \MgIII, leads to levels involved in a specific transition that could be more populated or depopulated, changing the resulting profile.

\subsection{New set of \texorpdfstring{$\Upsilon_{ij}$}{ECS}} \label{subsec:ECS_populations}

Models 1401b and 1401c differ in the effective collision strength $(\Upsilon_{ij})$ data. As it was detailed in Section \ref{subsec:ecs}, model 1401c included new calculations using the DW method for transition whose upper level is in the range between 26 (57\,262.76 $cm^{-1}$) and 54 (59\,430.517 $cm^{-1}$).

The method of calculating the $\Upsilon_{ij}$ values in both models, also produced a redistribution of \MgI energy levels, and among the ionized species \MgI, \MgII, and \MgIII between them. However, these changes were more moderate than those found in the previous section. Figure \ref{fig:1401c_1401b} shows the ratio of the population of each level between models, integrated over all the atmospheric heights. The population of the first levels of model 1401c is higher than that of model 1401b. Although this behavior is similar to the obtained in the previous section, the increase of the lower-lying level populations due to the DW $\Upsilon_{ij}$ data is one order of magnitude smaller. 

The overall results in the calculated spectra, obtained with this significant change in $\Upsilon_{ij}$ data, could be separated into various distinctive regions. 
For transitions within 1747 {\AA} and 12\,510 {\AA}, we have not found noticeable differences driven by it, although at the end of this region a small change begins to emerge. Figure \ref{fig:VIS+NIR} illustrates these findings since the calculated spectral lines from models 1401b and 1401c are superposed in each case. In these examples, and across the mentioned range, models of 85 levels (1401b and 1401c) produce almost identical line profiles, indicating that e+\MgI collisions have not been played an important role in the line formation. It was not possible to find solar observations to compare from 12\,510 to 22\,500 {\AA}, although the calculated spectra from both models presented an almost negligible difference of around 2 to 4\% in some lines. The same pattern was observed until approximately 30\,000 {\AA}.
MIR lines with wavelengths higher than that value, were more sensitive to changes of $\Upsilon_{ij}$ data. We found differences larger than 6\% at the center of several lines.

Figure \ref{fig:IR} shows several examples in which model 1401b produced deeper lines than model 1401c. Although synthetic lines are still deeper than observations in both cases, the CCC + DW + SEA\&VRM combination for $\Upsilon_{ij}$ (model 1401c) produced an important improvement in matching the observations throughout the IR range considered. In this spectral range, we found two kinds of improvement in the calculated spectra with 1401c over 1401b. On one side, a better match with observations in all the transitions with new DW $\Upsilon_{ij}$ data. On the other, we also found an improvement in transitions involving the first 25 levels with the same $\Upsilon_{ij}$ in both models.

To analyze the effect in line formation caused by the different sets of $\Upsilon_{ij}$ data used in the calculation, we selected three lines to use as proxies of the changes observed in the calculated spectra. These lines are: the insensitive NUV feature at 2\,780.6 {\AA} (shown in Fig. \ref{fig:2780}), and the improved MIR features 33\,200 {\AA} and 71\,097 {\AA} (shown in Figs. \ref{fig:IR}c and \ref{fig:IR}d respectively). We computed the formation region of these lines in the atmosphere, using the \textit{contribution function} (or attenuated emissivity; \cite{fontenla:2007}). Atmospheric parameters such as temperature, pressure, and density at the formation height give information about the conditions of the plasma in which the NLTE populations and the emerging line profile were calculated.

Figures \ref{fig:ECS_contrib}a and \ref{fig:ECS_contrib}c present the contribution functions for the transitions that form the features 2\,780.6 {\AA} and 71\,097 {\AA}, respectively. The plot for 33\,200 {\AA} is not shown because it is very similar to 71\,097 {\AA}.

The center of the 2\,780.6 {\AA} feature was formed in two distinctive regions. The former was located near the temperature minimum region, at the end of the photosphere. And the rest, at the beginning of the chromospheric plateau (see the atmospheric regions in Fig. \ref{fig:1401}). The peaks of the contribution function were at 666 km and 737 km, and around 1000 km, respectively. This wide range of heights in the atmosphere corresponds to temperatures between $\sim$4000 to 6000 K. 

The formation region was within $\sim$175--390 km for the 71\,097 {\AA} feature. The maximum was located in the middle of the photosphere at 250 km, with temperatures from $\sim$4500 to 5000 K. The 33\,200 {\AA} feature was produced within the same range, from 180 to 280 km.
Both lines were formed in a narrower temperature range and deeper in the atmosphere than the 2\,780.6 {\AA} feature. These examples illustrate spectral features formed at different heights, where the atmospheric conditions in each case could constrain the physical processes that populate the energy levels involved. Our results suggest that lines which are formed in low-density atmospheric regions could be less affected by changes in the data that describe collisions with electrons.

Figure \ref{fig:ECS_contrib}b shows a comparison of the $\Upsilon_{ij}$ parameters used in the 1401b and 1401c models for the line 2\,780.6 {\AA}, at the temperature of the formation region. The $\Upsilon_{ij}$ value used at a specific temperature was obtained interpolating these data and then used to compute the NLTE population. Although the $\Upsilon_{ij}$ calculated using the DW method (adopted in model 1401c) is, on average, eight times higher than the values computed using SEA\&VRM method (adopted in model 1401b), the synthetic spectral feature calculated for each model remained unchanged (see in Fig. \ref{fig:2780}).

Instead, the 71\,097 {\AA} feature was formed at photospheric heights, where the density of \MgI is more than two orders of magnitude higher than in the chromospheric plateau.
Figure \ref{fig:ECS_contrib}d shows that the $\Upsilon_{ij}$ data calculated using DW method is about three times higher than that computed using SEA\&VRM method. In this case, collisions with electron appear to be more relevant since a remarkable improvement in matching the ACE Solar Atlas spectral lines was reached using the 1401c model, compared with that computed using 1401b model (see Fig. \ref{fig:IR}d).

The 33\,200 {\AA} feature is formed by a transition between levels 13 (54\,192.256 $cm^{-1}$) and 24 (57\,204.228 $cm^{-1}$), with CCC $\Upsilon_{ij}$ data in both models. This fact indicates that we had to find another way to study the influence of the $\Upsilon_{ij}$ values over the spectral line formation that takes into account the coupled system.

As the depth of a spectral line is related to the difference in the population of the levels involved, our results point out that there was a change among the energy level population distribution of \MgI due to the inclusion of DW calculations replacing SEA\&VRM data. It provides an alternative way to analyze the influence of the $\Upsilon_{ij}$ on the calculated line. We used the \textit{level population ratio} calculated as $N_{low}/N_{up}$, where $N_{low}$ or $N_{up}$ refers to the populations of each level involved in the transition at the formation height. This ratio allowed to detect changes in the level populations that could be enough to affect the depth of the related spectral line.

Figure \ref{fig:levelsratio} presents the \textit{level population ratio} for transitions 2\,780.6 {\AA} and 71\,097 {\AA}. The former resulted almost identical for both 1401b and 1401c models, as we expected. This result again illustrates that even when changes in the $\Upsilon_{ij}$ data (shown in Fig. \ref{fig:ECS_contrib}b) could be large, it has no influence in their \textit{level population ratio} and therefore neither in the spectral line. Collisional process could not be the dominant process for this transition. Regarding the 71\,097 {\AA} feature in which the collisional process may play an important role (Fig. \ref{fig:ECS_contrib}d), the \textit{level population ratio} is different for each model as we predicted, being higher for model 1401b than 1401c. 

Figure \ref{fig:ratio_33200} shows the \textit{level population ratio} for 33\,200 {\AA} feature. Although the $\Upsilon_{ij}$ for this profile is the same data calculated with CCC method in both models, it behaved like the \textit{level populations ratio} for 71\,097 {\AA} profile. 

\begin{figure}
\resizebox{\hsize}{!}{\includegraphics{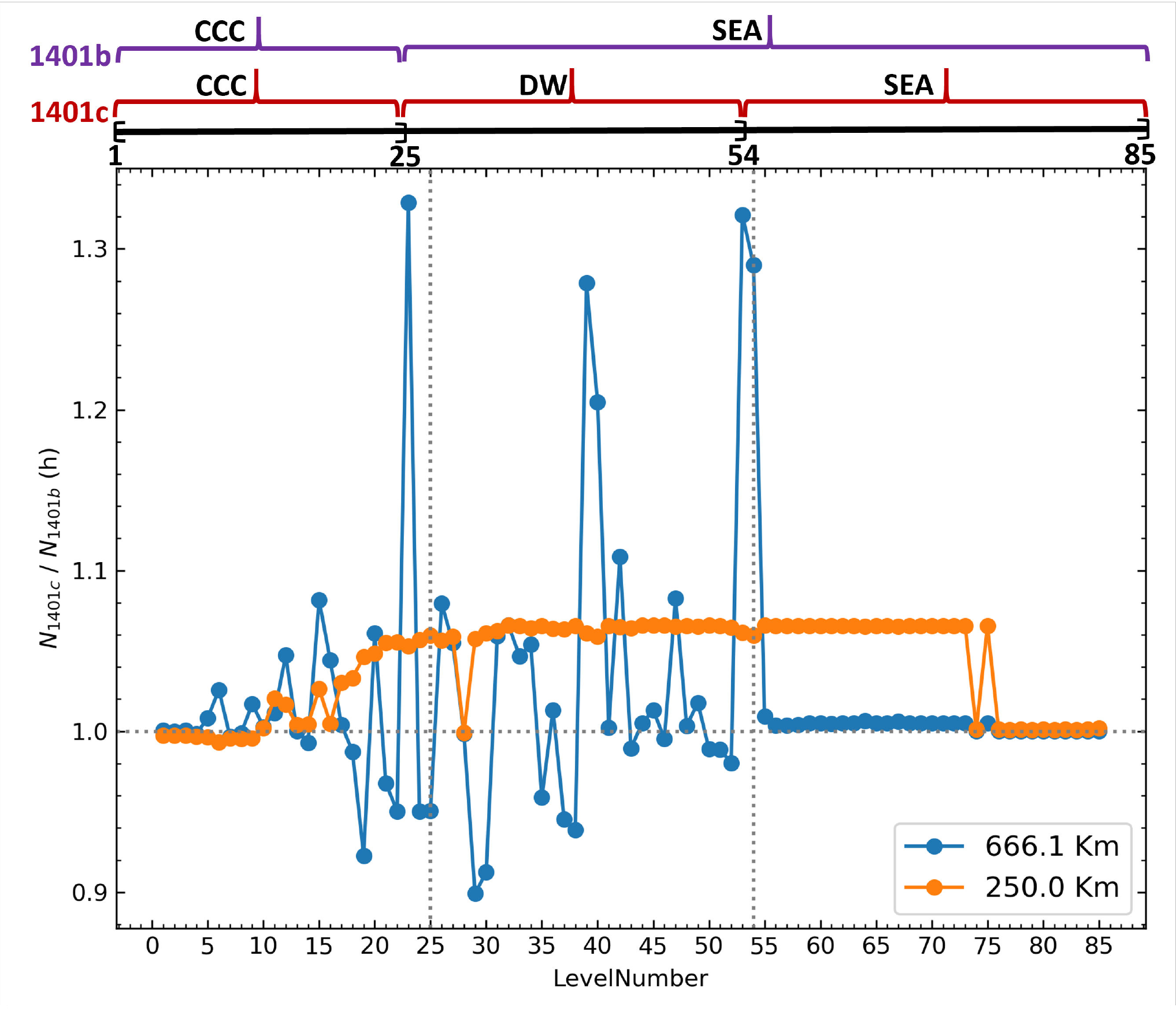}}
\caption{\textit{Model population ratio} as a function of the energy level, at 250 km (orange connected dots) and 666.1 km (blue connected dots) in the atmosphere. In the \textit{top} of the figure, it is marked the $\Upsilon_{ij}$ data used in each model. The \textit{upper} level determine the $\Upsilon_{ij}$ data used. e.g: for transition 24--38 (71\,097 {\AA}), the SEA\&VRM data are used in model 1401b (in purple) and DW data in model 1401c (in red).}
\label{fig:mods_ratio}
\end{figure}

To analyze the redistribution of population among energy levels at the formation height, we computed the \textit{model population ratio} $(N_{1401c}/N_{1401b})$ for the 85 energy levels, at several heights in the atmosphere. In the height range from $\sim$175 to 550 km, the population of model 1401c is larger than model 1401b for levels higher than level 20 ($5d\,^1D$, 56\,308.381 $cm^{-1}$). This ratio reaches a maximum of about 12\% at 350 km. Figure \ref{fig:mods_ratio} shows the \textit{model population ratio} at two representative formation heights, for feature 2\,780.6 {\AA} at 666.1 km, and features 71\,097 {\AA} and 33\,200 {\AA} at approximately 250 km.

Fig. \ref{fig:mods_ratio} shows that at 250 km in the photosphere, the \textit{model population ratio} for the first levels are slightly lower than 1. The population of model 1401c is redistributed to levels higher than $\sim$20, where the \textit{model population ratio} is bigger than unity in a $\sim$7\%. The reason for this redistribution is due to the DW $\Upsilon_{ij}$ are larger than SEA\&VRM $\Upsilon_{ij}$ data. The system is coupled through the statistical equilibrium equations, responding as a whole to the change of $\Upsilon_{ij}$ collisional parameters. The population of each level is affected by the rest of the levels present in the atomic model.

Regarding our proxy features, the 2\,780.6 {\AA} profile is formed by transitions from level 2  (21\,850.405 $cm^{-1}$) to 28  (57\,812.77 $cm^{-1}$) at 666.1 km. The \textit{model population ratio} for both levels is close to 1 resulting in almost identical line profiles for models 1401b and 1401c at formation height, as Fig. \ref{fig:2780} indicates.

In the case of 71\,097 {\AA} feature, the \textit{model population ratio} for the levels involved (24 | 57\,204.228 $cm^{-1}$ and 38 | 59\,430.517 $cm^{-1}$) at 250 km, departes from unity $\sim$6\%, and $\sim$7\% respectively, as shows Fig. \ref{fig:mods_ratio}. This small difference of $\sim$1\% at the formation height is responsible for the change on the calculated line profile between models. 

The 33\,200 {\AA} feature is a transition between levels 13 (54\,192.256 $cm^{-1}$) and 24 (57\,204.228 $cm^{-1}$) at 250 km. The upper level involved in this transition has a higher population in model 1401c than in model 1401b, while the lower level has approximated the same population in both models. The extra \MgI population of level 24 in model 1401c, could come from higher levels whose DW $\Upsilon_{ij}$ are higher than that of SEA\&VRM in model 1401b. Although the numerical value of the $\Upsilon_{ij}$ in this specific transition is the same in both models, the population changes due to the value of $\Upsilon_{ij}$ of the rest of the levels. 

In general, when the \textit{model population ratio} at the formation height for each level involved in a transition is unity, then the transition will have the same line profile in the calculated spectra for both models. On the contrary, if the \textit{model population ratio} of the lower level is less than that of the upper level, then the line profile will be shallower in model 1401c. As Fig. \ref{fig:mods_ratio} indicates, all line profiles with formation height of $\sim$250 km (where most of the MIR line formed) whose lower level is approximately less than $\sim$20 and the upper level is bigger than $\sim$20, except for level 28 (57\,812.77 $cm^{-1}$), will be shallower in model 1401c, improving the match with solar observations.

\begin{figure}
\centering
\resizebox{\hsize}{!}{\includegraphics{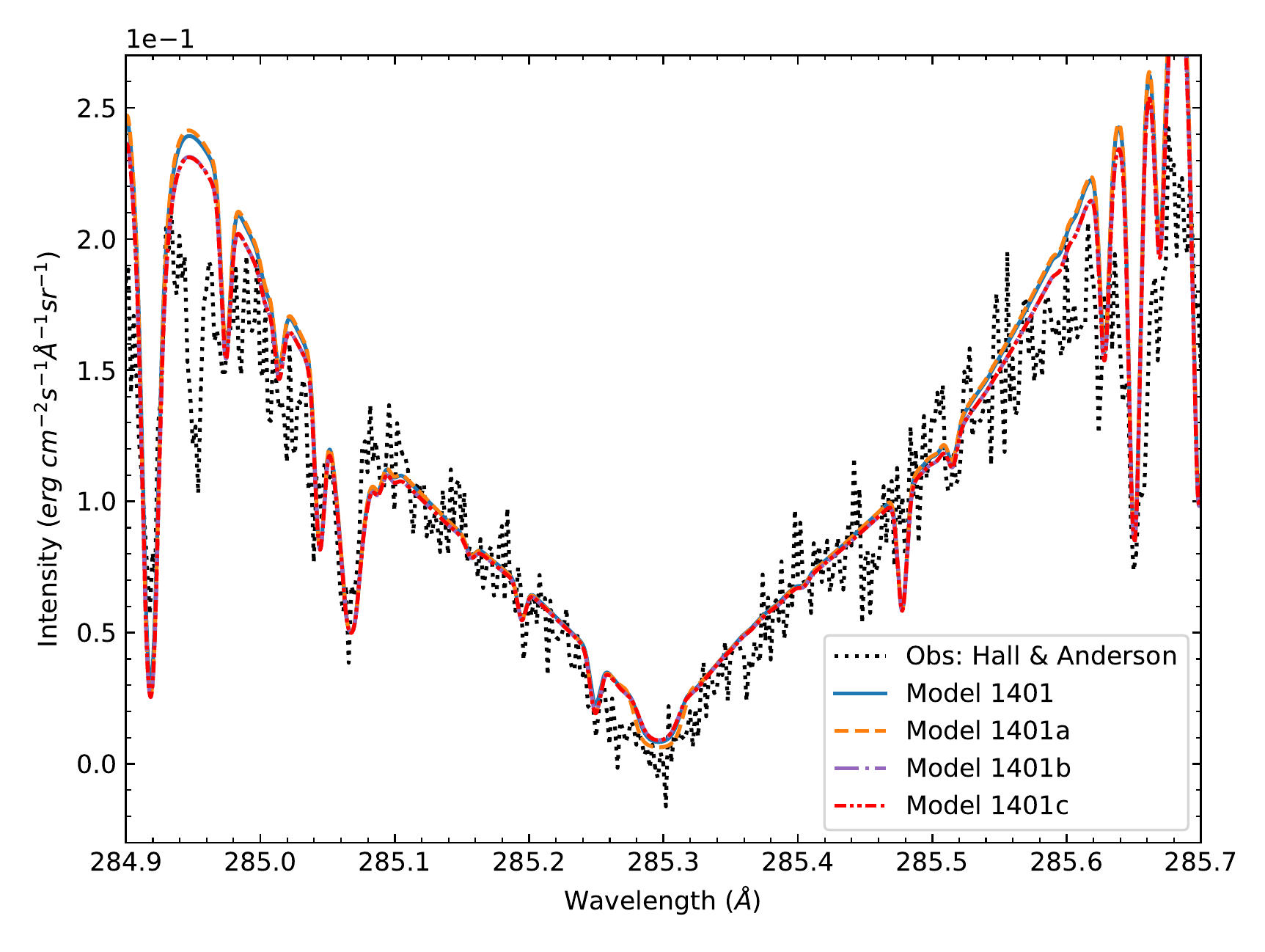}}
\caption{Comparison of the H\&A solar irradiance observations of the 2852 {\AA} line profile with the synthetic line calculated from models 1401, 1401a, 1401b, and 1401c.}
\label{fig:mgnuv}
\end{figure}

One significant \MgI feature in the NUV range is the resonance 2852 {\AA} line profile resulting from the transition between levels 1 and 3 (35\,051.264 $cm^{-1}$). This profile could be used as an important diagnostic of the thermal structure in solar or stellar chromospheres. Figure \ref{fig:mgnuv} shows a comparison of the H\&A solar irradiance observations with the synthetic line calculated with models 1401, 1401a, 1401b, and 1401c. The calculated profiles are lightly sensitive to an increase of the number of levels since 1401 is overlapped with 1401a. The same behavior occurs with 1401b and 1401c, although in this case the center of the line is almost negligible shallower and the upper part of the wings are lower than the other models.

\cite{fontenla:2016} and \cite{tilipman:2020} built atmospheric models for the dM stars GJ 832 and GJ 581 using the stellar version of the \textit{SRPM}. These stellar models were calculated with our starting \MgI atomic model (1401). The NUV synthetic spectra computed in these works revealed an incorrect emission in the center of the 2852 {\AA} line, in comparison with STIS G230L \textit{HST} observations.
In a forthcoming paper, we will address this problem, testing our new \MgI atomic model 1401c, given the different plasma characteristics of the atmosphere of stars cooler than our Sun.

\section{Conclusions} \label{sec:conclusions}

The selected data of the \MgI atomic model used in solving the NLTE population and calculating the spectra are crucial factors to improve the agreement of \MgI line profiles in solar observations from NUV to MIR.

Updates in the oscillator strengths, and the broadening parameter describing radiative, Stark, and van der Waals processes produced noticeable changes in the calculated line profiles. Our model 1401a showed a clear improvement over the initial model 1401, driven by a redistribution of Mg population among its ionization states and mainly throughout the \MgI atomic levels itself.

To reach our goal of including important features of \MgI from the NUV to the MIR spectral range, we developed a new level structure shown by the Grotrian diagram of Fig. \ref{fig:grotriam} for the models 1401b and 1401c. Starting from a sketchy atomic structure of 26 levels, we ended with a robust structure of 85 levels capable of reproducing a long list of transitions, many of which have never been calculated in NLTE and studied in detail before. 

Regarding the effective collision strength $\Upsilon_{ij}$ parameter, we made use of the latest CCC calculation by \cite{barklem:2017} for the first 25 levels. For the following levels, we considered the widely used formulation of SEA\&VRM (yielding the model 1401b). To find a better agreement in the line profiles, we performed a new calculation for these parameters from levels 26 (57\,262.76 $cm^{-1}$) to 54 (59\,430.517 $cm^{-1}$) with the multi-configuration Breit–Pauli distorted–wave (DW) method (used by model 1401c). This method has seldom been used in stellar astrophysics.  
We found an additional improvement to that obtained updating and expanding the level structure due to the DW $\Upsilon_{ij}$ calculation. The synthetic spectra obtained using the model 1401c reached a better agreement with solar observations from NUV to MIR (1747--71\,099 {\AA}) than the other models considered. Nevertheless, the observed profiles are still shallower than the calculated lines.

These results open new challenges for future work. We will continue improving the calculation of the $\Upsilon_{ij}$ data from level 55 to 85 using the DW method. The inclusion of this new data in the statistical equilibrium equations could produce a redistribution of the Mg population similar to that obtained in the present work. It could imply even shallower line profiles than those obtained with the model 1401c, potentially achieving an excellent match with solar observations. 

The DW method has shown to be a better alternative to SEA\&VRM for calculating the $\Upsilon_{ij}$ parameters of \MgI when close-coupling methods are impracticable or data are not available. We suggest that its application to other atomic species should be a more reliable option to semi-empirical calculations.  

As the effective collision strength parameters depend on the temperature, the plasma characteristics of the region in which a transition is formed determine the NLTE population and the resulting line profile. Forthcoming work will mainly cover the NLTE \MgI line formation in stars with different spectral types, with a special interest in stars cooler than our Sun with planets in its habitable zone.

\begin{acknowledgements} \label{sec:acknow}
We wish to sincerely thank to Dr. Jeffrey Linsky for his detailed revision of our manuscript since it has helped us improve our work.
This work has made use of the \textit{VALD} database, operated at Uppsala University, the Institute of Astronomy RAS in Moscow, and the University of Vienna.
We acknowledge to 1995 Atomic Line Data (R.L. Kurucz and B. Bell) Kurucz CD-ROM No. 23. Cambridge, Mass.: Smithsonian Astrophysical Observatory.

\end{acknowledgements}

\bibliographystyle{aa_url} 
\bibliography{aa} 

\end{document}